\documentclass[prd,twocolumn,showpacs,reprint,preprintnumbers,nofootinbib,superscriptaddress]{revtex4-1}

\RequirePackage[colorlinks=true
,urlcolor=blue
,anchorcolor=blue
,citecolor=blue
,filecolor=blue
,linkcolor=blue
,menucolor=blue
,linktocpage=true
,pdfproducer=medialab
,pdfa=true
]{hyperref}

\usepackage{overpic}
\usepackage{amsmath,amssymb,amsthm,amsfonts}
\usepackage{graphicx,tabularx}
\usepackage{color}
\usepackage{multirow}
\usepackage{comment}
\usepackage{enumitem}
\usepackage{cleveref}
\usepackage{slashed}
\usepackage[normalem]{ulem}
\usepackage{scalerel}
\usepackage{mathtools}
\usepackage{fontawesome5}

\setlist[description]{leftmargin=0.3cm}
\setlist[itemize]{leftmargin=0.5cm}

\newcommand{\be}{\begin{equation} \begin{aligned}}
\newcommand{\ee}{\end{aligned} \end{equation}}
\newcommand{\beqa}{\begin{eqnarray}}
\newcommand{\eeqa}{\end{eqnarray}}

\def\figureautorefname~#1\null{Fig.\,#1\null}
\def\tableautorefname~#1\null{Tab.\,#1\null}

\def\equationautorefname~#1\null{Eq.\,(#1)\null}

\newcommand{\mev}{\text{MeV}}
\newcommand{\gev}{\text{GeV}}
\newcommand{\tev}{\text{TeV}}

\newcommand{\m}{\text{m}}

\newcommand{\mrad}{\text{mrad}}

\renewcommand{\eqref}[1]{Eq.~(\ref{eq:#1})}

\RequirePackage[normalem]{ulem}

\definecolor{greenST}{RGB}{0,201,87}

\crefname{section}{Sec.}{Secs.}
\crefname{figure}{Fig.}{Figs.}
\crefname{equation}{Eq.}{Eqs.}
\crefname{appendix}{Appendix}{Appendices}
\crefname{table}{Table}{Tables}

\begin{document}

\preprint{DESY-22-084, UCI-HEP-TR-2022-06}

\title{FLArE up dark sectors with EM form factors at the LHC Forward Physics Facility}

\author{Felix Kling}
\email{felix.kling@desy.de}
\affiliation{Deutsches Elektronen-Synchrotron DESY, Notkestr. 85, 22607 Hamburg, Germany}

\author{Jui-Lin Kuo}
\email{juilink1@uci.edu}
\affiliation{Department of Physics and Astronomy,  University of California, Irvine, CA 92697-4575, USA}

\author{Sebastian Trojanowski}
\email{strojanowski@camk.edu.pl}
\affiliation{Astrocent, Nicolaus Copernicus Astronomical Center Polish Academy of Sciences, ul.~Rektorska 4, 00-614, Warsaw, Poland}
\affiliation{National Centre for Nuclear Research, ul.~Pasteura 7, 02-093 Warsaw, Poland}

\author{Yu-Dai Tsai}
\email{yt444@cornell.edu}
\email{yudait1@uci.edu}
\affiliation{Department of Physics and Astronomy,  University of California, Irvine, CA 92697-4575, USA}

\begin{abstract}
Despite being mostly secluded, dark sector particles may feebly interact with photons via a small mass-dimension 4 millicharge, a mass-dimension 5 magnetic and electric dipole moment, or a mass-dimension 6 anapole moment and charge radius. If sufficiently light, the LHC may produce an intense and collimated beam of these particles in the far forward direction. We study the prospects of searching for such dark sector particles with electromagnetic form factors via their electron scattering signature in the Forward Liquid Argon Experiment (FLArE) detector at the Forward Physics Facility (FPF). We find that FLArE can provide new probes of sub-GeV dark particles with dipole moments and strong sensitivities for millicharged particles in the 100 MeV to 100 GeV region. This complements other search strategies using scintillation signatures or dark matter direct detection and allows for probing strongly interacting dark matter motivated by the EDGES anomaly. Along with the FORMOSA detector, this leads to a very diverse and leading experimental program in the search for millicharged particles in the FPF.
\end{abstract}

\maketitle

\section{Introduction}
\label{sec:intro}

Shedding light on beyond the Standard Model (BSM) constituents of nature remains of primary importance for our understanding of elementary interactions. Of particular interest are possible couplings of the new species to the Standard Model (SM) photons that could leave an impact on the entire cosmological history, as well as could nowadays be probed experimentally. In particular, while dark matter (DM) species are generically expected to have suppressed such interactions, even very weak couplings of this type could lead to striking observational consequences. Therefore, pushing the boundaries of our understanding of BSM couplings to photons helps us understand how the dark sector of the Universe works.

Such couplings are generally expected to occur in many BSM scenarios, either directly or at a loop-induced level. At low energies, the resulting couplings can be written in terms of effective field theory (EFT) operators, which generate electromagnetic (EM) form factors of the dark species. This can lead to both electrically charged and neutral new particles with suppressed interactions with the SM photons that can be within reach of current and future searches.

In particular, the existence of BSM particles with small EM charges can be postulated, which is also well-motivated as an indirect test of Grand Unified Theory and string theory~\cite{Pati:1973uk, Georgi:1974my, Wen:1985qj, Shiu:2013wxa}. As a minimal scenario, such millicharged particles (mCPs) can arise from a small hypercharge BSM coupling, or through the kinetic mixing between a dark photon and the SM photon~\cite{Holdom:1985ag}. Interestingly, mCPs can have a non-negligible relic abundance and make up a fraction of DM, thus leaving their impact on cosmology, direct detection experiments, and indirect detection  searches~\cite{Davidson:2000hf, Pospelov:2007mp, Vogel:2013raa}.  When the scattering cross-section of millicharged DM is larger than a certain critical value, its flux can be attenuated from interactions with the atmosphere and earth crust before reaching the terrestrial direct detection detectors and thus escape the direct detection experiments~\cite{Rich:1987st,Mahdawi:2018euy,Emken:2019tni,Foroughi-Abari:2020qar}. Dedicated experiments~\cite{Rich:1987st,Emken:2019tni} were conducted and proposed to search for such ``strongly interacting'' DM. In addition, the interaction between mCPs and baryons around the 21 cm epoch can explain the recent EDGES observation~\cite{Bowman:2018yin,Munoz:2018pzp}. While later investigations show that the parameter space of the minimal scenario is strongly constrained~\cite{Fialkov:2018xre,Berlin:2018sjs,Barkana:2018qrx,Barkana:2018lgd,Slatyer:2018aqg,Boddy:2018wzy,Kovetz:2018zan,Creque-Sarbinowski:2019mcm}, more complicated scenarios can be invoked to reconcile the anomaly~\cite{Liu:2019knx}. These DM candidates can be meaningfully constrained and probed by existing and future accelerator experiments as well as large neutrino observatories~\cite{Davidson:2000hf,CMS:2012xi,CMS:2013czn,Jaeckel:2012yz,Essig:2013lka,Haas:2014dda,Ball:2016zrp,milliQan:2021lne,Ball:2020dnx,Liu:2018jdi,Prinz:1998ua,Berlin:2018bsc,Gninenko:2018ter,Arefyeva:2022eba,Magill:2018tbb,Kelly:2018brz,Beacham:2019nyx,Harnik:2019zee,ArgoNeuT:2019ckq,Foroughi-Abari:2020qar,Marocco:2020dqu}.

On the other hand, even if dark states are EM neutral, higher-dimensional effective couplings to the SM photons can still be present. Specifically, for a Dirac dark state $\chi$, one can study magnetic dipole moments (MDM) and electric dipole moments (EDM) at mass-dimension $5$, and anapole moment (AM) and charge radius interaction (CR) at mass-dimension $6$~\cite{Kavanagh:2018xeh}. Notably, dark sector dipole moments have similar structures and potential common origin to the neutrino dipole portals (see, e.g.,~\cite{Magill:2018jla,Shoemaker:2020kji,Brdar:2020quo,Dasgupta:2021fpn}), cf. recent studies about searches at the LHC~\cite{Jodlowski:2020vhr,Ismail:2021dyp}. Assuming the above effective interaction terms, the dark state can make up the thermal DM relic through the standard freeze-out mechanism, and be probed in direct and indirect searches~\cite{Pospelov:2000bq,Sigurdson:2004zp,Schmidt:2012yg,Ho:2012bg,Kopp:2014tsa,Ibarra:2015fqa,Sandick:2016zut,Kavanagh:2018xeh,Chu:2018qrm,Trickle:2019ovy,Arina:2020mxo}.

The BSM species coupled to the SM photons are also studied independently of the DM motivation and are searched for in numerous experiments. In general, $\chi$ particles with EM form factors can be produced by interactions associated with photons; therefore, accelerator-based experiments and stars are ideal sites to study these scenarios. In particular, colliders~\cite{Davidson:2000hf,CMS:2012xi,CMS:2013czn,Jaeckel:2012yz,Essig:2013lka,Haas:2014dda,Ball:2016zrp,Liu:2018jdi,Foroughi-Abari:2020qar,Ball:2020dnx,milliQan:2021lne}, proton-fixed target and neutrino facilities~\cite{Magill:2018tbb,Harnik:2019zee,Kelly:2018brz,Beacham:2019nyx,ArgoNeuT:2019ckq,Marocco:2020dqu}, as well as lepton facilities~\cite{Prinz:1998ua,Berlin:2018bsc,Gninenko:2018ter,Arefyeva:2022eba} provide strong current constraints and good future discovery prospects for the mCP model and higher-dimensional operators~\cite{Banerjee:2017hhz,Akesson:2018vlm,Battaglieri:2016ggd,Aubert:2001tu,Abe:2010gxa,Chu:2020ysb,Kuo:2021mtp}. Particles with small EM charges can be produced from atmospheric collisions of cosmic rays and be observed by large neutrino observatories~\cite{Plestid:2020kdm,ArguellesDelgado:2021lek}. SM precision observables (including lepton $g-2$ and the fine structure constant measurements at different energies), as well as astrophysical and cosmological signatures, can also be used to probe higher-dimensional operators~\cite{Chu:2018qrm,Chu:2019rok,Chang:2019xva}.

The far-forward region of the LHC presents new exciting opportunities to conduct novel searches of such BSM species. At this unique location, even distant and small detectors can search for new light unstable particles~\cite{Feng:2017uoz, Feng:2017vli, Kling:2018wct, Feng:2018noy}, as well as study high-energy neutrinos~\cite{FASER:2021mtu, Ismail:2020yqc, Mosel:2022tqc, Kelly:2021mcd}. This observation lead to the approval of FASER~\cite{FASER:2018ceo, FASER:2018eoc, FASER:2018bac, FASER:2021cpr, FASER:2021ljd, Boyd:2803084}, FASER$\nu$~\cite{FASER:2019dxq, FASER:2020gpr}, and SND@LHC~\cite{SHiP:2020sos, Ahdida:2750060} detectors to take data during LHC Run 3, as well as to the proposal of a dedicated Forward Physics Facility (FPF)~\cite{MammenAbraham:2020hex, Anchordoqui:2021ghd, Feng:2022inv} for the High Luminosity LHC (HL-LHC) era. The FPF would host several experiments, including the proposed FORMOSA detector~\cite{Foroughi-Abari:2020qar} in which mCPs could be searched for via a low ionization signal. Compared to other mCP studies, this search benefits from the TeV-scale center-of-mass (CM) energy of $pp$ collisions at the LHC such that mCPs up to the mass of order few tens of GeV can be probed with $\mathcal{O}(10^{-3})$ EM charge. The focus on the far-forward direction allows for maximizing the flux of mCPs such that a relatively compact detector can provide strong bounds on this scenario in the future.

In this paper, we discuss alternative detection signatures based on scatterings of high-energy dark particles with EM form factors producing detectable soft electron recoils in the FPF detectors. To this end, we present our results, for the recently proposed Forward Liquid Argon Experiment (FLArE) experiment~\cite{Batell:2021blf} which is envisioned to employ liquid argon time projection chamber (LArTPC) technology to detect visible signals from such scatterings, and to study neutrino interactions~\cite{Anchordoqui:2021ghd, Feng:2022inv}. Given the excellent capabilities of LArTPC detectors to study soft electron-induced signals at the energy level of above $\mathcal{O}(10~\mev)$, and to reconstruct such events, FLArE will be well-suited to study low-energy scattering signatures. We also stress, however, that similar sensitivity could be expected for other FPF scattering experiments, i.e., Advanced SND@LHC and FASER$\nu$2 detectors, cf. Refs~\cite{Anchordoqui:2021ghd, Feng:2022inv} for further discussion, provided their final design allows for sensitive searches for low-energy electron recoils.

As shown below, the combined research agenda of the FPF will lead to a variety of experimental approaches and world-leading detection prospects in the search for mCPs in a wide range of their masses, with potentially important connections to the aforementioned astrophysical observations. We also present, for the first time, expected FPF bounds for the higher-dimensional operators. In particular, we show the projected bounds for electric and magnetic dipole moment that can improve current constraints for dark states in the sub-GeV mass range.

This paper is organized as follows. In~\cref{sec:int_EMff}, we describe the considered particle model. In~\cref{sec:production}, we discuss dominant dark state production channels, followed by detection strategy and background estimation in~\cref{sec:signature}. In~\cref{sec:FLArE}, we demonstrate the projected sensitivity of FLArE and compare it with existing constraints. Finally, we draw a conclusion in~\cref{sec:discussion}.

\section{Dark states with electromagnetic form factors}
\label{sec:int_EMff}

We consider a scenario in which the SM photon is the new physics mediator which couples to dark sector particles $\chi$ carrying EM form factors. To have the most affluent set of EM form factors, in this work we take $\chi$ to be a Dirac particle. Starting from the lowest dimensions and till mass-dimension $6$ in the EFT framework,\footnote{At mass-dimension $7$, one can have Rayleigh operators describing interactions between $\chi$ and two photons~\cite{Kavanagh:2018xeh}, which is beyond the scope of this work and will not be considered further.} allowed effective operators are mass-dimension $4$ electric monopole, mass-dimension $5$ magnetic/electric dipole moment (MDM/EDM) and mass-dimension $6$ anapole moment (AM) and charge radius (CR). 
The Lagrangian describing interactions between $\chi$ and the SM photon can be expressed as~\cite{Chu:2018qrm},
\be
\!\!\mathcal{L}_\chi & \supset \epsilon e \bar\chi \gamma^\mu \chi A_\mu + \dfrac{1}{2} \mu_\chi \bar{\chi} \sigma^{\mu\nu} \! \chi F_{\mu\nu} + \dfrac{i}{2} d_\chi \bar{\chi} \sigma^{\mu\nu} \!\gamma^5 \!\chi F_{\mu\nu} \!\! \\
&\ \  - a_\chi \bar{\chi} \gamma^\mu \gamma^5 \chi \partial^\nu F_{\mu\nu} + b_\chi \bar{\chi} \gamma^\mu \chi \partial^\nu F_{\mu\nu}\,,
\label{eq:interaction_L}
\ee
where $F_{\mu\nu}$ is the field strength of the SM photon, $\epsilon$ is the fraction of $\chi$'s charge relative to the elementary charge $e$, $\mu_\chi$ and $d_\chi$ are coefficients of MDM and EDM with dimension $[M]^{-1}$, $a_\chi$ and $b_\chi$ are coefficients of AM and CR with dimension $[M]^{-2}$, and $\sigma^{\mu\nu} \equiv i[\gamma^\mu, \gamma^\nu]/2$.

Below the electroweak scale, the effective couplings to photons in \cref{eq:interaction_L} can be induced from more general hypercharge couplings. At mass-dimension 4, the tiny millicharge can generally arise from a small hypercharge $\epsilon'e$ carried by $\chi$ before the electroweak symmetry breaking. The interaction Lagrangian between millicharged $\chi$ and the SM ${\rm U}(1)_Y$ gauge field $B_\mu$ can be expressed as $\epsilon'e \bar{\chi} \gamma^\mu \chi B_\mu$. This term can be added to explicitly break the charge quantization, or can be more theoretically motivated and source from a kinetic mixing between a dark photon and the SM photon~\cite{Holdom:1985ag}. After electroweak symmetry breaking, we then observe that $\epsilon = \epsilon'\cos\theta_W$ with $\theta_W$ being the weak mixing angle. Similarly, the coupling between $\chi$ and the $Z$ boson will be induced. We discuss its impact on the sensitivity reach of the FPF experiments for the mCP model in \cref{sec:FLArE}.

\begin{figure*}[t]
\centering
\includegraphics[width=0.43\textwidth]{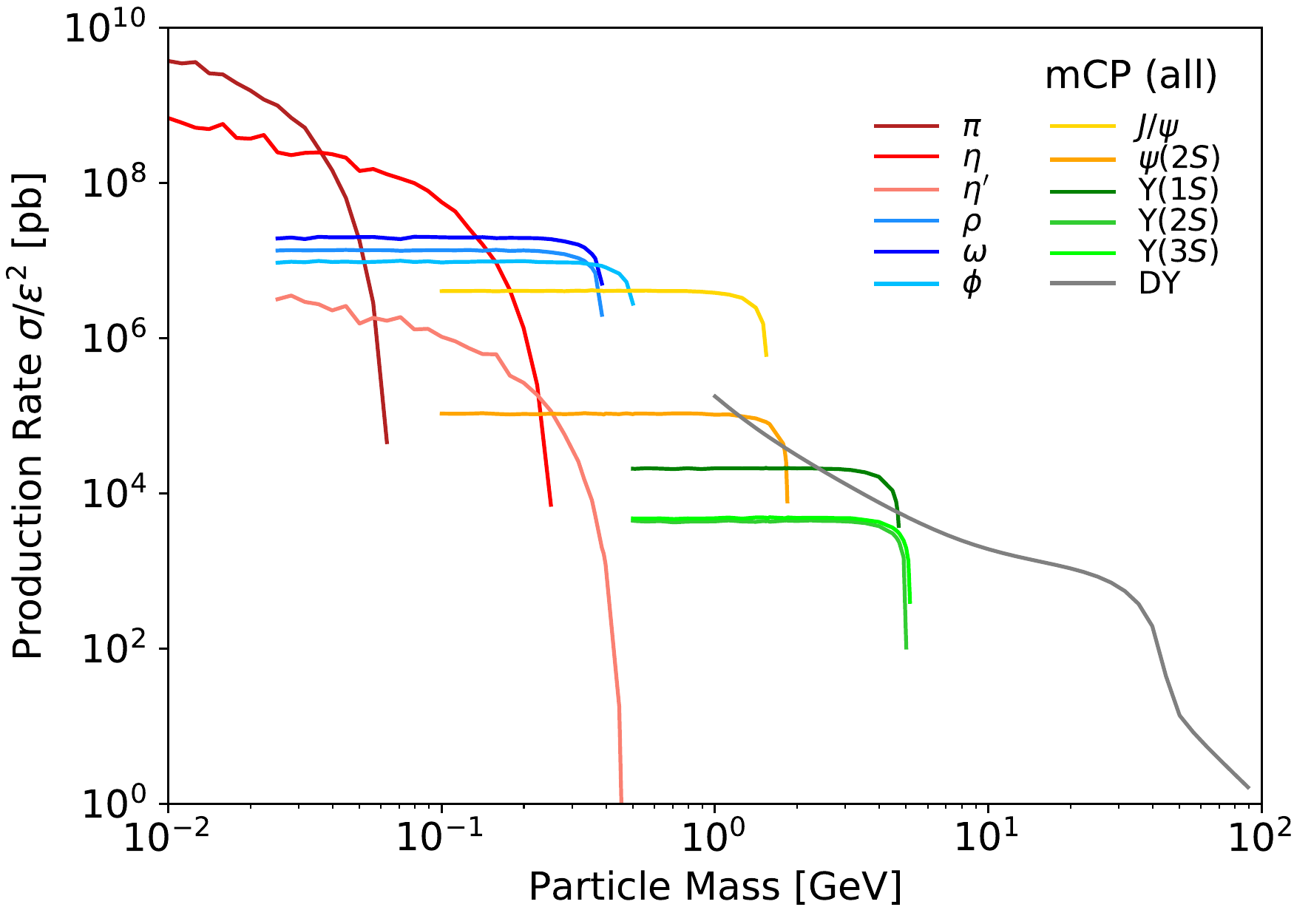}
\includegraphics[width=0.43\textwidth]{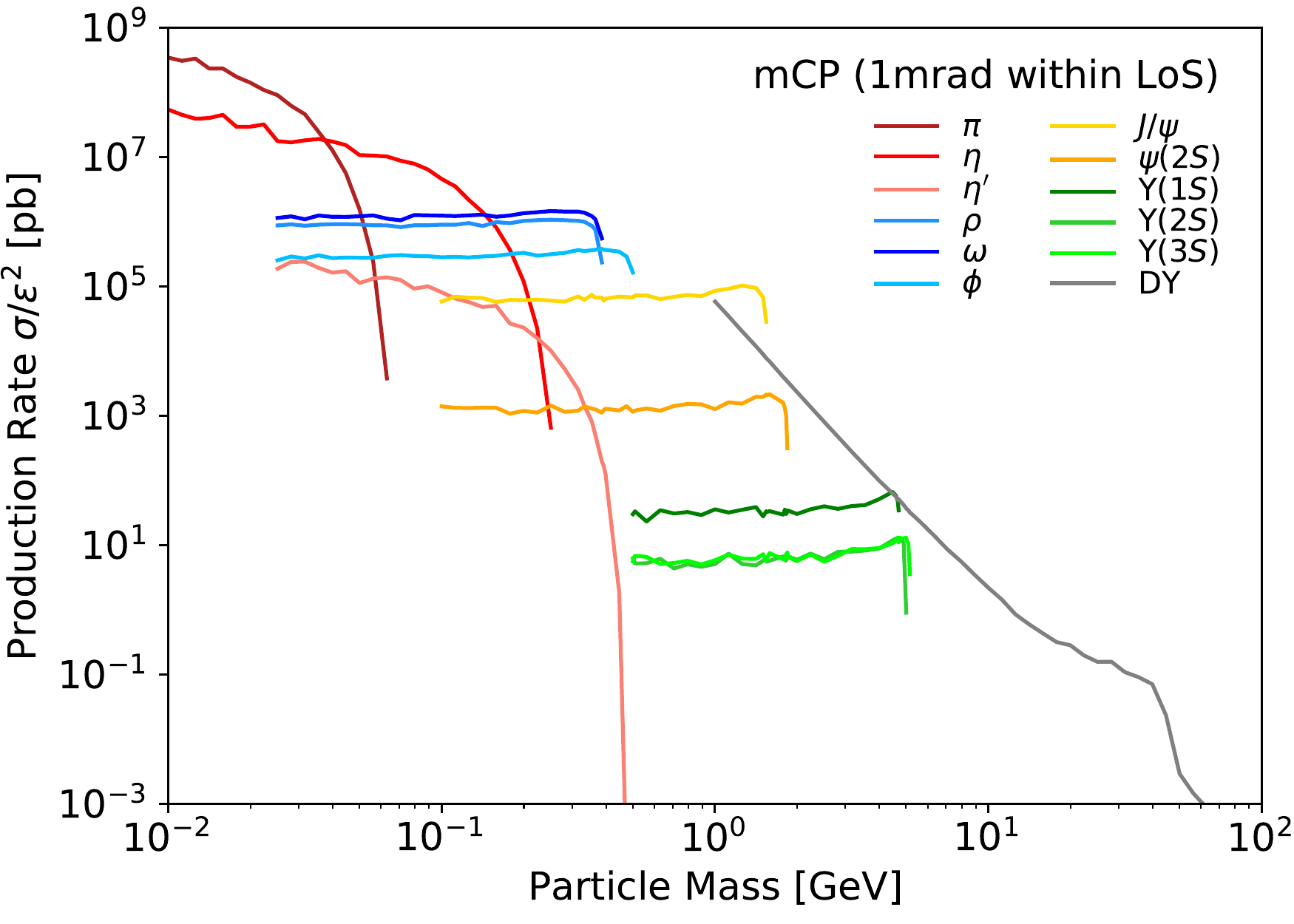}
\includegraphics[width=0.43\textwidth]{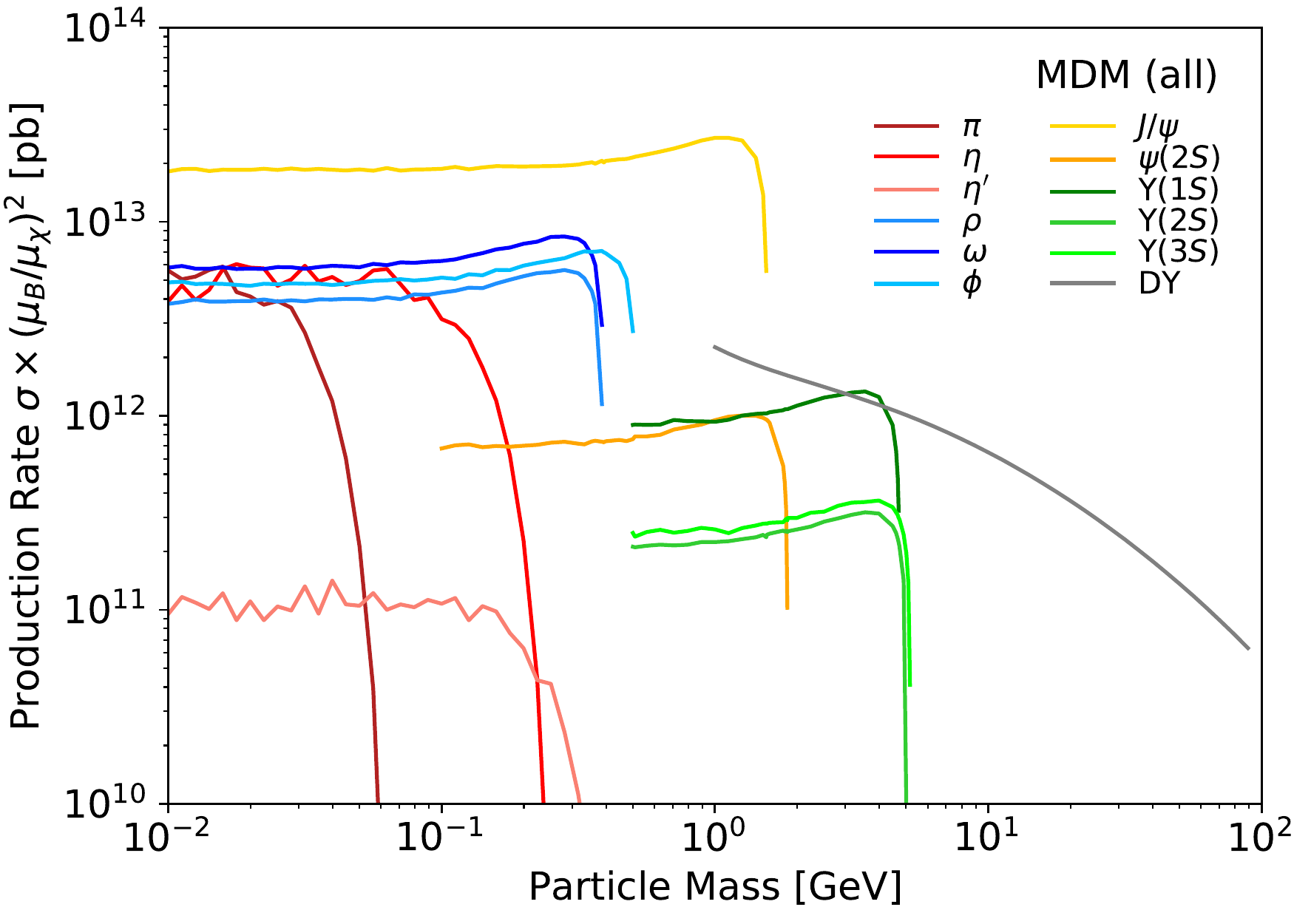}
\includegraphics[width=0.43\textwidth]{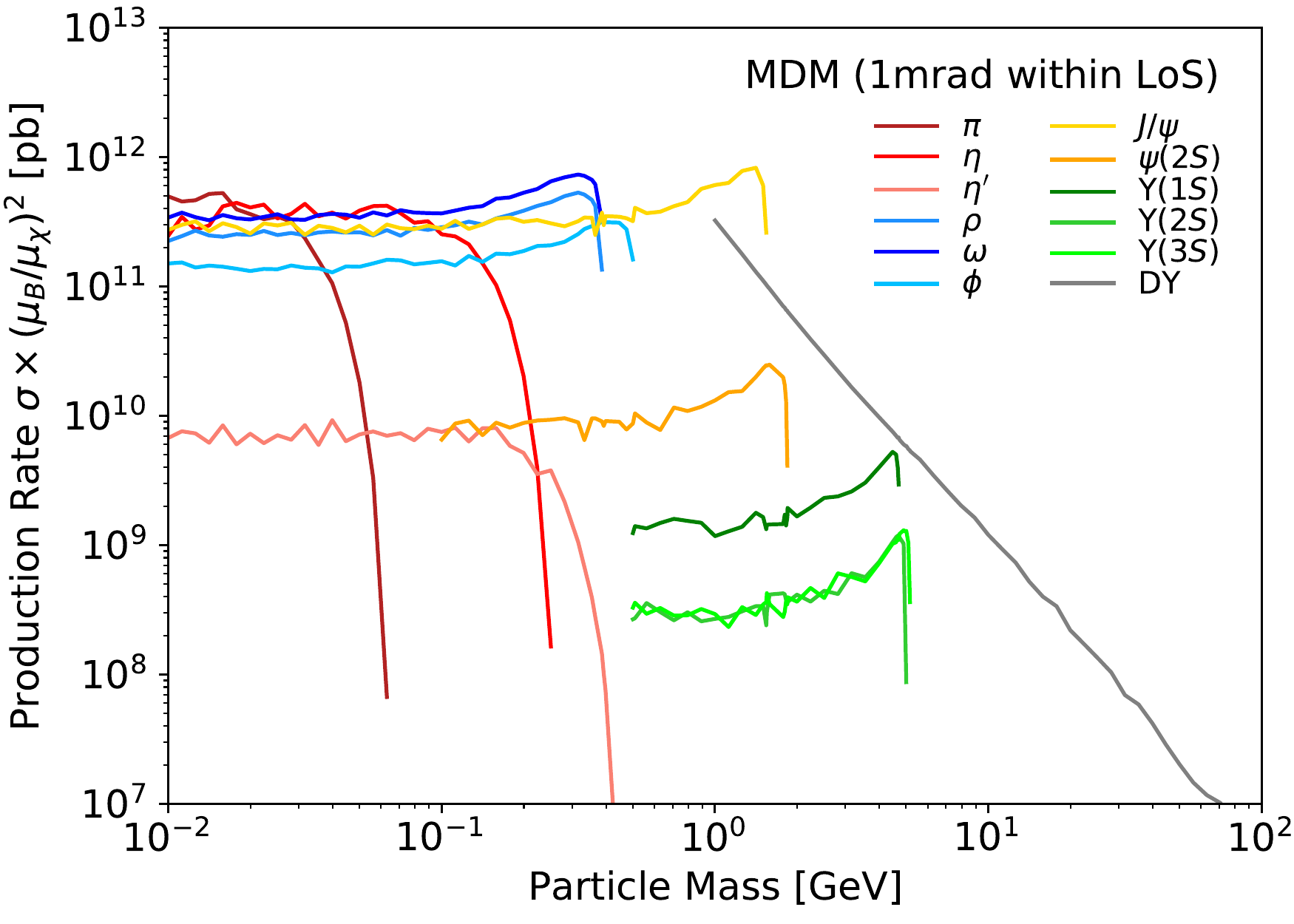}
\includegraphics[width=0.43\textwidth]{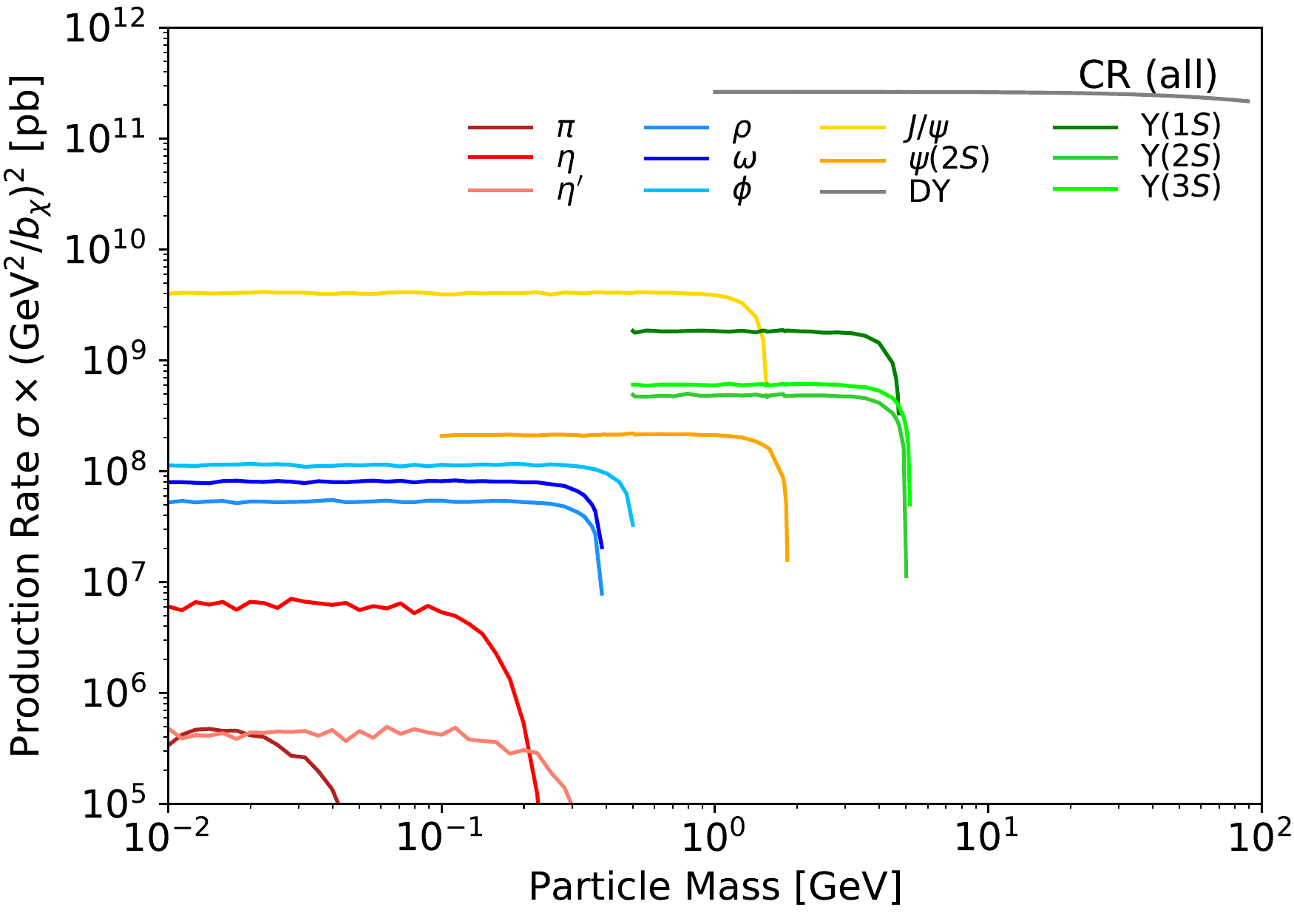}
\includegraphics[width=0.43\textwidth]{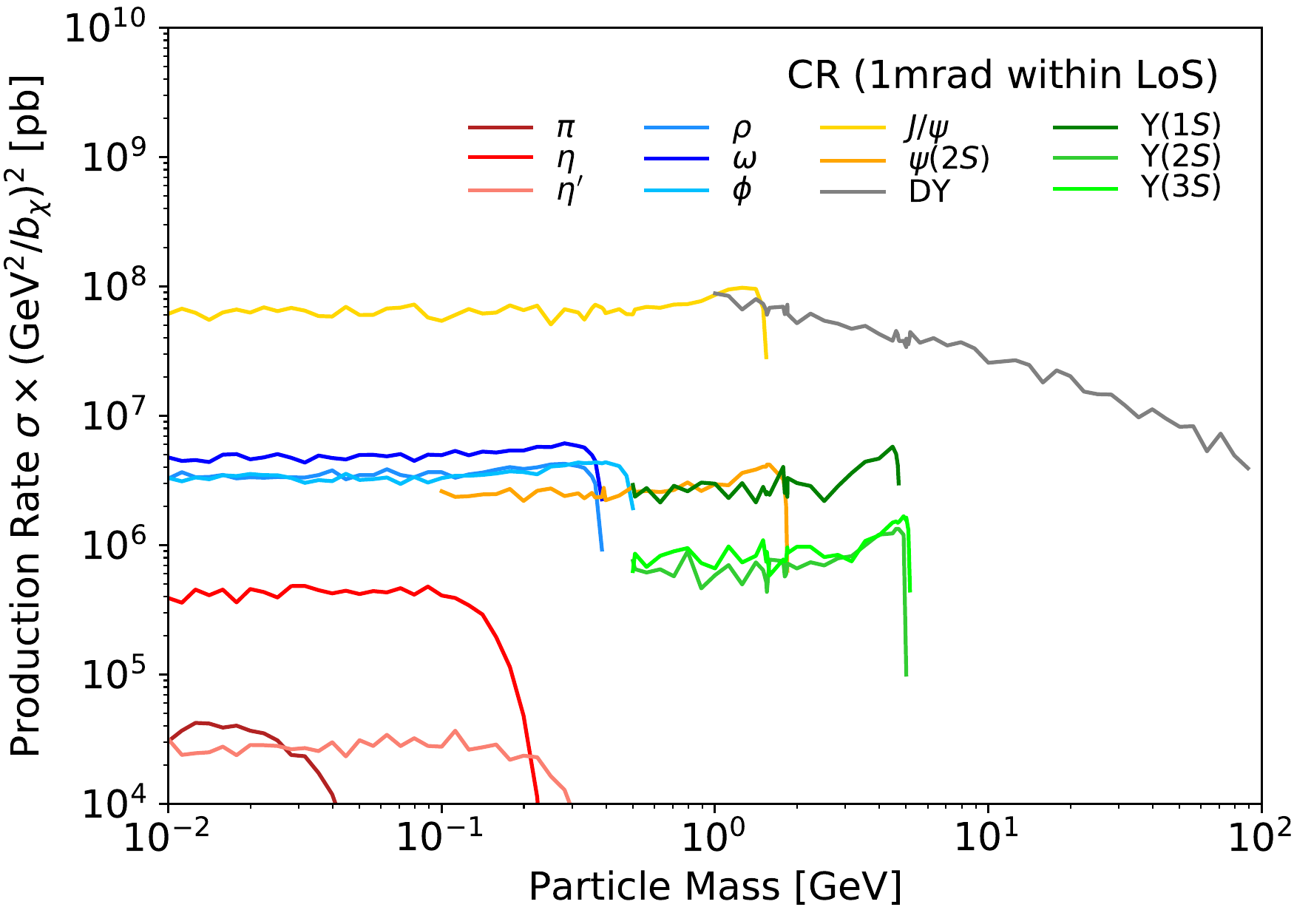}
\caption{Production rates of dark states with millicharge (top panels) and EM form factors with MDM (center) and CR (bottom) within the whole forward hemisphere (left panels) and within $1~\mrad$ of the Line of Sight (LoS) (shown on the right). Due to the nature of effective operators, production from heavier mesons and the Drell-Yan (DY) process becomes increasingly important for higher mass-dimension, in contrast to the mCP case where production from $\pi^0$ decay dominates; see main texts for further details. Note that production rates of EDM (AM) case only differs with that of MDM (CR) at the kinematic endpoints, thus we do not show them here.}
\label{fig:prod}
\end{figure*}

In the case of higher-dimensional operators, similar issues arise from replacing $F_{\mu\nu}$ with the hypercharge field strength tensor $B_{\mu\nu}$ in the operators in \cref{eq:interaction_L}. Notably, this has important consequences for both the missing energy searches for $\chi$s at the LHC and its DM phenomenology for $m_\chi\gtrsim 100~\gev$, as in both cases one probes regions of the parameter space beyond the EFT validity regime for the couplings to $F_{\mu\nu}$~\cite{Arina:2020mxo}. Instead, at lower energy scales characteristic for light hadron decays, the expected sensitivity is driven by the photon couplings. We then find it sufficient to focus on this regime below, while we also comment on consequences of employing the hypercharge couplings. 

Once the hypercharge coupling is taken into account, the interaction is \textit{a priori} UV-completed for mass-dimension $4$. In contrast, for mass-dimension 5 and 6 operators, EFT is not valid anymore when we start to probe the full particle spectrum of the underlying theory. The effective theory considered in this work can be UV-completed by considering, e.g., compositeness of $\chi$~\cite{Foadi:2008qv,Bagnasco:1993st,Antipin:2015xia} or loop contribution via new electric-charged states at UV-scale~\cite{Raby:1987ga,Pospelov:2008qx}. Assuming the UV-scale is not within the reach of FPF, we adopt the effective Lagrangian~\cref{eq:interaction_L} in the following calculations.

\section{Production Channels at the Forward Physics Facility}
\label{sec:production}

The LHC is the most energetic particle collider build thus far and one of its primary objects for its remaining operations is the search for new heavy particles at the TeV scale. These particles are expected to decay roughly isotropically, and large experiments have been build around the LHC interaction points to detect their decay products. As first pointed out in Ref.~\cite{Feng:2017uoz}, the LHC could also be used to probe a different class of new particles which are light and weakly coupled particles. This idea utilizes that the LHC produces an enormous number of hadrons, which are mainly produced in the direction of the colliding beams and typically escape undetected through the beam pipe. If kinematically allowed, these hadrons could undergo rare decays into new light weakly coupled states and produce an energetic and strongly collimated beam of these particles in the forward direction. In this study, we perform a dedicated Monte Carlo simulation using \texttt{FORESEE}~\cite{Kling:2021fwx} to estimate the flux of dark sector particles $\chi$ produced at the LHC. 

If sufficiently light, $\chi$ can be produced through both three-body decays of pseudoscalar mesons $P \to \gamma \chi\bar\chi$ and two-body decays of vector mesons $V \to \chi\bar\chi$. In the following, we consider the pseudoscalar mesons $P = \pi^0, \eta, \eta'$ as well as the vector mesons $V=\omega, \rho, \phi, J/\Psi, \Psi(2S), \Upsilon(nS)$ as parent particles and use particle spectra provided by \texttt{FORESEE}~\cite{Kling:2021fwx}. Here, the spectra of light mesons were obtained using the \texttt{EPOS-LHC}~\cite{Pierog:2013ria} event generator as implemented in the \texttt{CRMC}~\cite{CRMC} interface. For the heavy Charmonium and Bottomonium states, we use the spectra presented in Ref.~\cite{Foroughi-Abari:2020qar} which were obtained using \texttt{Pythia~8}~\cite{Sjostrand:2006za, Sjostrand:2014zea} and tuned to LHCb data~\cite{Aaij:2015rla, Aaij:2018pfp, Aaij:2019wfo}. 

Heavier $\chi$, for which the above-mentioned meson decays are kinematically forbidden, are primarily produced via  the Drell-Yan process $q\bar{q} \to \chi\bar\chi$. We simulate this using \texttt{MadGraph~5}~\cite{Alwall:2014hca} with the model file adapted from Ref.~\cite{Chu:2020ysb}. To ensure that the parton distribution functions are well defined, we only consider this partonic scattering production mode for masses $m_\chi > 1~\gev$. For smaller masses, and as a result also smaller momentum transfers, the partonic picture loses its validity. In this case, other processes such as production via coherent Bremsstrahlung off the proton beam could play a role~\cite{Brem_precite}.

In \cref{fig:prod} we present the production rates of the dark sector state $\chi$ at the LHC with 14~TeV center-of-mass energy for the different production channels mentioned above. The left panels correspond to the total production rate, while the right panels show the production rate after requiring the particle to be within 1~mrad around the beam collision axis, which roughly corresponds to the angular size of FLArE. 

For mCPs, the production rate is roughly energy-independent. Thus $\chi$-production is dominated by decay of lighter mesons due to their larger fluxes. However, the hierarchy is different for higher-dimensional operators for which production rate is energy-dependent. For example in the case of two-body vector meson decays, their decay branching fractions scale as $\text{BR}_{V\to\chi\bar\chi}\sim 1$ for mCPs, $\text{BR}_{V\to\chi\bar\chi}\sim M^2$ for EDM/MDM and $\text{BR}_{V\to\chi\bar\chi}\sim M^4$ for CR/AM, where $M$ is the vector meson mass (see \cref{app:meson_decay} for more details). As a result, the importance of heavier meson decays or the Drell-Yan (DY) production increases for dark sector states with higher dimensional operators. After imposing the angular cut, we find that for mass-dimension $5$ operators (MDM and EDM) production from $J/\psi$ decay can be comparable to lighter meson decay, while for mass-dimension $6$ operators (AM and CR) production from $J/\psi$ decay and the DY processes remain dominant throughout the considered $m_\chi$-range. Notably, while the DY production can probe large center-of-mass energies in the hard collision, for which mass-dimension $5$ and $6$ operators with only an $F_{\mu\nu}$ coupling could suffer from the aforementioned lack of the EFT validity, we will see below that this production process does not determine the FLArE sensitivity reach in the currently allowed regions of the parameter space of these models.

In the case of the mCP model, we also recognize in the upper panels of \cref{fig:prod} the impact of the $Z$ boson coupling induced by small hypercharge. This can be seen as a characteristic peak of the total DY production rate for $m_\chi \simeq m_Z/2$, where the effective mCP production via the $Z$ resonance occurs. We expect similar peaks to be present for other models if the hypercharge coupling was taken into account (not shown in the plots). This, however, does not affect lighter $\chi$s for which the dominant production is through hadron decays.

\section{Signatures at FLArE}
\label{sec:signature}

Stable dark sector particles which are produced in the far-forward region of the LHC can be studied via their scatterings in the detectors in the FPF. In our discussion, we focus on electron-recoil signals induced by new physics species. These can be more efficiently disentangled from neutrino-induced backgrounds than signatures based on scatterings off nuclei~\cite{Batell:2021aja, Batell:2021snh}, especially in searches targeting low energy depositions in the detector. In the following, we discuss several strategies to maximize the sensitivity reach and we comment on expected backgrounds.  

\subsection{Signal}

\textit{Scattering a-la DM signal -} The first signature that we consider resembles the single electron-recoil signature, which has previously been considered for a light DM search in FLArE~\cite{Batell:2021blf}. In the relativistic regime, we find for the scattering cross section of the process $\chi e \to \chi e$ that $d\sigma/ dE_R \propto E_R^x$ where $E_R$ is the electron recoil energy and $x = -2, -1, 0$ for mCP, MDM/EDM and AM/CR, respectively; see \cref{eq:scattering_xsec}. Therefore we can infer that for lower mass-dimension operators, the event rate is enhanced for softer recoils, which is especially important for mCP searches~\cite{Foroughi-Abari:2020qar}; see also the following discussion. 

In our analysis below, we assume the following energy thresholds for the recoiled electron
\begin{equation}
30~\textrm{MeV}\,(300~\textrm{MeV}) \lesssim E_{R,\textrm{single}} \lesssim 1~\textrm{GeV},
\label{eq:singlecuts}
\end{equation}
where the lower cut of $30~\textrm{MeV}$ corresponds to the assumed LArTPC detector capabilities to study soft electron tracks, cf. Refs~\cite{DUNE:2015lol}. In order to illustrate the impact of this cut on our analysis, we also present below the expected FLArE sensitivity reach in the search for mCPs assuming an increased lower energy threshold which is set at $300~\textrm{MeV}$. The upper energy threshold of $1~\gev$ allows for suppressing neutrino-induced backgrounds while maintaining the signal rate, as discussed below. 

\textit{Double-hit with softer recoils -} Since the differential scattering cross section of mCPs is more IR-biased, further improvement in the sensitivity reach can be expected for an even lower energy threshold in LArTPC detector~\cite{Magill:2018tbb}. While lowering it too much could lead to additional backgrounds, these can be circumvented by focusing on double-hit events, in which the signature consists of two coincident and collinear $\chi$ scatterings off electrons~\cite{Harnik:2019zee}. We, therefore, present the expected sensitivity of FLArE to such a search for mCPs assuming the following cuts on both electron recoil energies
\begin{equation}
5~\textrm{MeV} \lesssim E_{R,\textrm{double}} \lesssim 1~\textrm{GeV},
\end{equation}
where the lower energy threshold to detect each of the hits in argon is chosen following Ref.~\cite{DUNE:2015lol}. The mCP-induced signal, in this case, would consist of two simultaneous hits which, given the large boost factors and very small deflection angles of $\chi$s, define the line pointing towards the direction of the ATLAS IP.

\subsection{Background}

The experimental signature of our interest in FLArE consists of a single scattered electron or two coincident such electron recoil signals. The signatures of this kind can also be mimicked by other types of effects. This leads to possible backgrounds that we briefly discuss below. We first focus on backgrounds induced by the two types of SM species which can reach the FPF experiments while being produced at the ATLAS IP, namely neutrinos and muons. We then comment on possible other sources of backgrounds that remain more difficult to estimate without detailed detector simulations.

\textit{Neutrino-induced backgrounds -} Signatures of new physics particles scattering in the detector often resemble similar processes characteristic to the SM neutrinos. In particular, the signal consisting of single scattered electrons in FLArE has already been studied in the context of light DM searches in Ref.~\cite{Batell:2021blf}. It has been shown that for low visible energy depositions in the detector, $30~\textrm{MeV} \lesssim E_{R,\textrm{single,loose}}\lesssim 20~\textrm{GeV}$, and thanks to the use of additional angular cuts, the expected neutrino-induced background rate can be as low as $\mathcal{O}(20)$ events during the HL-LHC era in the 10-tonne FLArE detector placed in the FPF along the beam collision axis. The most significant background contribution is associated with the neutrino-electron scatterings, while quasi-elastic and resonant nuclear scatterings of $\nu_e$ also contribute non-negligibly. Instead, neutrino-induced deep inelastic scattering events can typically be rejected due to the presence of additional visible tracks in the detector. 

Notably, the number of neutrino-induced events expected in the FPF is highly suppressed in this low-energy regime. This is primarily dictated by the large energy of incident neutrinos produced in the far-forward region of the LHC, typically of order $E_\nu\sim 200-300~\textrm{GeV}$, see Refs~\cite{Bai:2020ukz, Kling:2021gos, Anchordoqui:2021ghd, Feng:2022inv, Bai:2021ira, Bai:2022jcs} for further discussion. Instead, low-energy neutrinos are produced more isotropically at the LHC. As a result, the remaining low-energy background from $\nu-e$ scatterings in FLArE is primarily associated with interactions of high-energy neutrinos with $E_\nu>100~\textrm{GeV}$ that can, occasionally, generate soft electron recoils.

We employ this fact in the analysis below by even further reducing the electron recoil energies required for our signal events in the single scattered electron signature, cf. \cref{eq:singlecuts}. In this case, we find less than $\mathcal{O}(1)$ expected neutrino-induced background events in FLArE, while the number of coincident double scattering events is significantly lower. In our estimates, we use far-forward neutrino fluxes and spectra presented in Ref.~\cite{Batell:2021aja} obtained using the fast neutrino flux simulation introduced in Ref.~\cite{Kling:2021gos} and we use \texttt{GENIE}~\cite{Andreopoulos:2009rq, Andreopoulos:2015wxa} to study neutrino interactions.

Neutrino-induced backgrounds can also appear due to $\nu$ scatterings in the rock and other material in front of FLArE. Such interactions can induce secondary particles and EM showers entering the detector from outside. Charged such species can be vetoed, similarly to the vetoing capabilities envisioned for FASER/FASER 2 detectors~\cite{FASER:2018ceo,FASER:2018bac}. Instead, rare events with secondary neutral hadrons entering the detector’s fiducial volume with no charged counterpart reaching FLArE would be more difficult to veto. Such events could mimic single scattered electron signatures by inducing additional photons in the detector, if these are energetic enough and if the remaining neutral hadrons remain undetected. We leave a detailed analysis of this background contribution for the future detector simulation. Since they could correspond to only a tiny fraction of neutrino interactions in front of the detector, we assume these background remains suppressed in the current analysis. 

\textit{Muon-induced backgrounds -} Instead, high-energy muons passing through the FPF detectors could more straightforwardly generate signals mimicking scattered electrons. This is primarily due to muon-induced photons produced in the detector~\cite{Batell:2021blf}. In order to suppress the rate of muons traveling through the FPF, the installation of a dedicated sweeper magnet has been proposed in the LHC tunnel which could deflect forward-going muons before they reach the experimental facility~\cite{Anchordoqui:2021ghd,Feng:2022inv}. In addition, such events are expected to be vetoed by detecting a charged muon entering FLArE. For this purpose, in order to improve event timing and triggering beyond the limitations of the drift time in LArTPC detectors, it is envisioned to equip FLArE with an additional light readout system~\cite{Feng:2022inv}. Muons could also generate secondary charged and neutral hadrons due to their interactions both inside and outside of the detector, cf. discussion in Ref.~\cite{FASER:2019dxq} for the FASER$\nu$ experiment. In the following, we assume that all such events are vetoed or rejected as not signal-like, as discussed above for neutrino-induced events.

\textit{Other background sources -} Further backgrounds corresponding to empty detector time frames with no neutrino scatterings in the detector can appear, e.g., due to ambient gamma-ray activity, intrinsic radioactivity, or electronic noise, cf. Ref.~\cite{ArgoNeuT:2018tvi} for discussion for the ArgoNeuT detector. In particular, these have been identified as important limitations of the mCP search in LAr detectors assuming order $\mev$ detection thresholds~\cite{Harnik:2019zee}. Instead, as discussed above for the single scattered electron signature, we employ the increased threshold of order $30~\mev$ or so, which leads to proper EM showers detectable in LArTPCs, cf. e.g., Ref.~\cite{DUNE:2015lol} for further discussion. In the case of the double-hit signature, a significant background reduction can be achieved even for lower energy thresholds $\sim 5~\mev$ thanks to the requirement that both hits are aligned~\cite{Harnik:2019zee}.
\medskip

\section{Projected sensitivity of FLArE}
\label{sec:FLArE}

We study the sensitivity reach for the FLArE detector placed in the FPF on the beam collision axis at a distance $L=620~\m$ away from the ATLAS IP. We assume the relevant detector geometry (fiducial volume)
\begin{equation}
\label{eq:detector_geometry}
\textrm{FLArE}:\ \ \ \Delta = 7\,\m,\ \ \ S_T = (1\,\m \times 1\,\m),
\end{equation}
where $\Delta$ is the detector length along the beam collision axis and $S_T$ is its transverse area. In our study, we primarily focus on the $10$-tonne FLArE detector, while for higher dimensional operators we also show how the expected reach could be improved for a larger $100$-tonne experiment FLArE-100 with $\Delta = 30~\textrm{m}$ and $S_T = (1.6~\textrm{m} \times 1.6~\textrm{m})$, cf. Ref.~\cite{Batell:2021blf} for a similar comparison for the DM search. At this stage, we assume $100\%$ signal detection efficiency for the dark states and study the electron scattering signatures discussed above. 

In the plots below, we present the result corresponding to $N_{\textrm{ev}} = 3$ expected new physics events in the detector. Assuming that backgrounds can be suppressed to a negligible level, this would correspond to sensitivity reach of FLArE in the proposed searches.

\subsection{Millicharged particles and dark matter}
\label{sec:sensitivity_mCP}

\begin{figure*}[t]
\centering
\includegraphics*[width=0.49\textwidth]{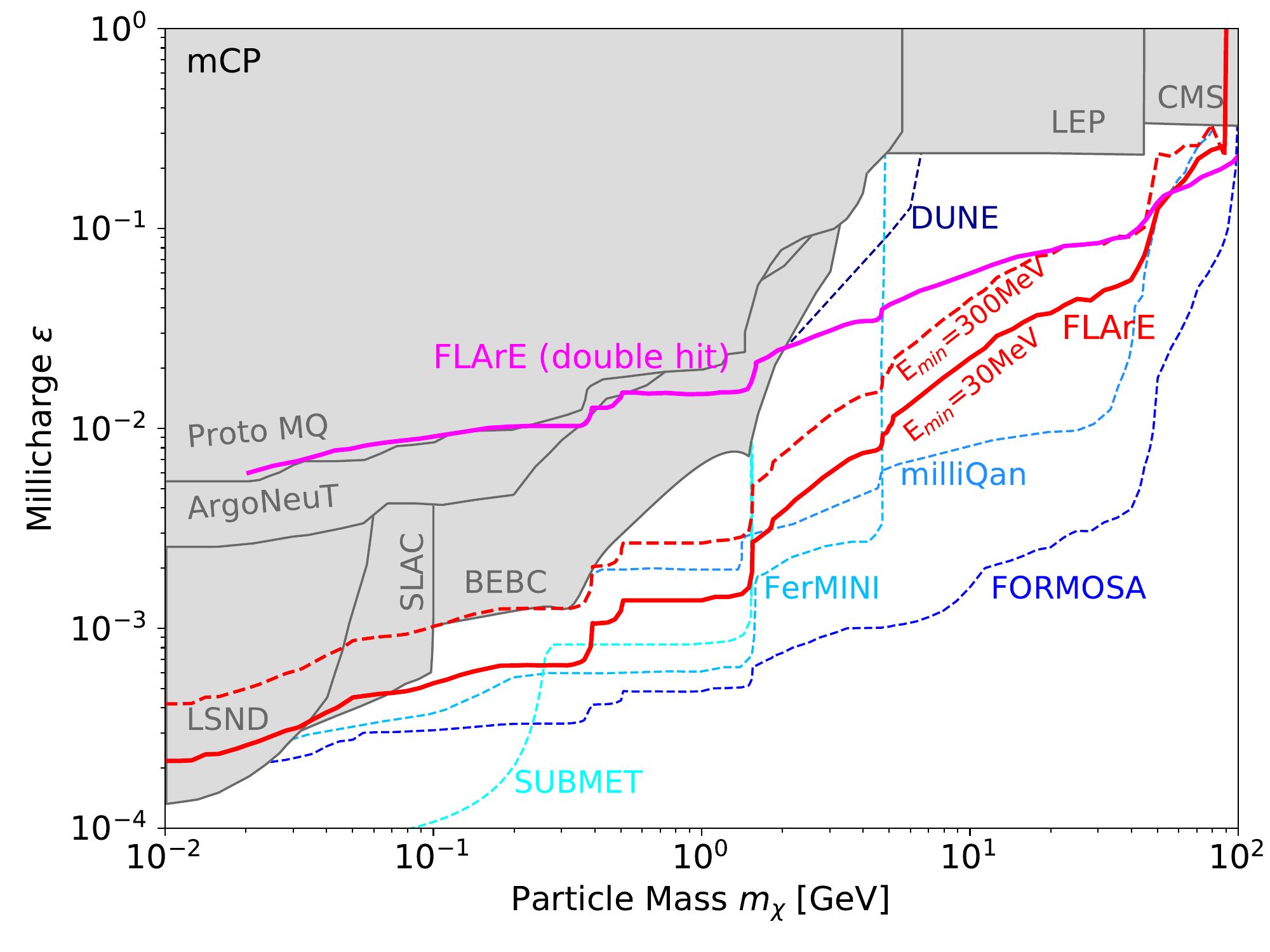}
\hfill
\includegraphics*[width=0.49\textwidth]{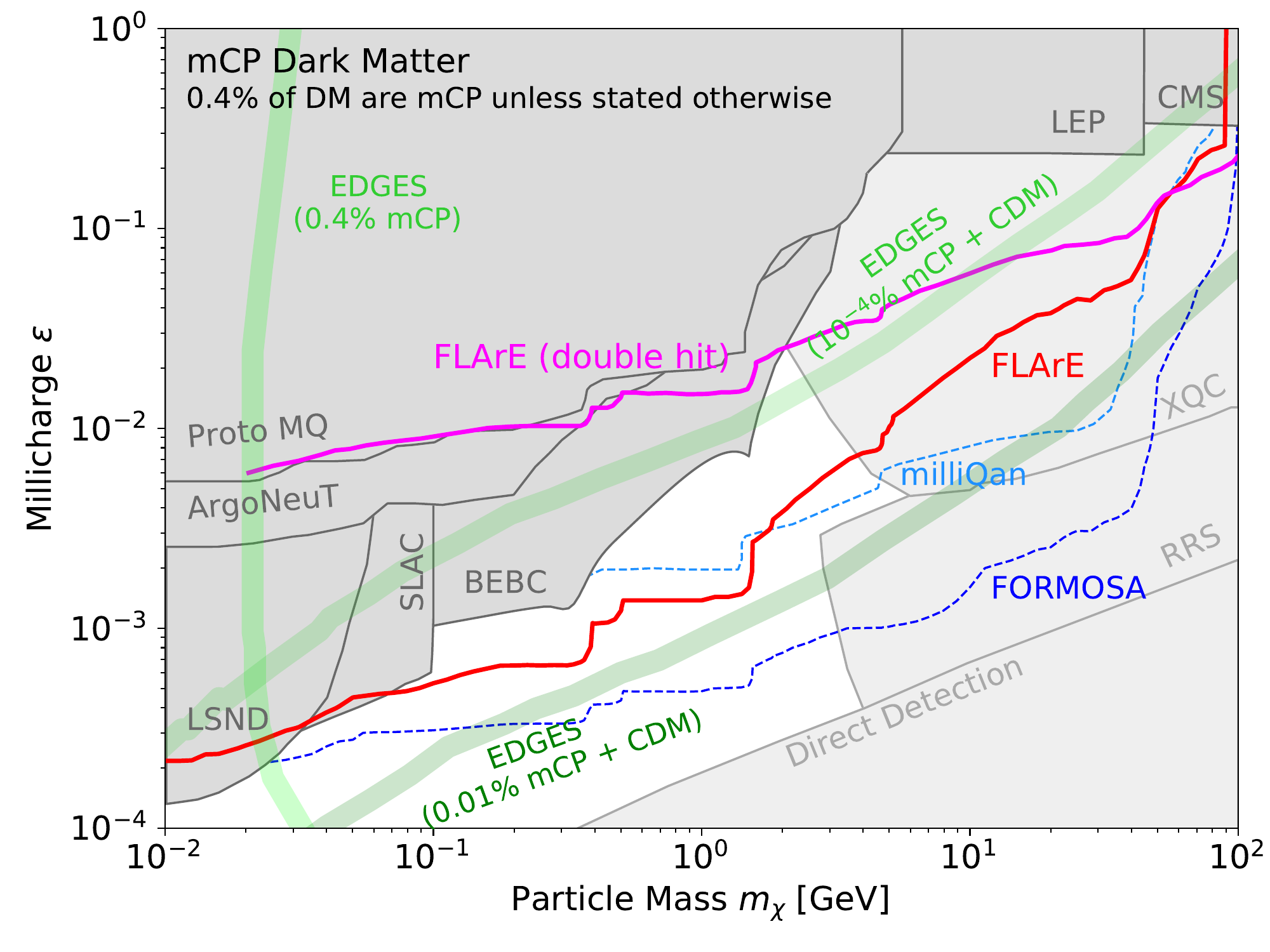}
\caption{FLArE reach using the signature similar to DM scattering ($30~\mev\lesssim E_R\lesssim 3~\gev$) and the one with a lower energy threshold ($E_R\gtrsim 1~\mev$) but requiring a double-hit event. The right panel shows the constraints for mCP dark matter, assuming $0.4\%$ of the dark matter being mCP. One green band ($0.4\%$) shows the parameter of mCP as dark matter to explain the EDGES anomaly, the other shows a model of $0.01\%$ dark matter being mCP, while the mCP interact with the rest of the cold DM to achieve additional cooling; see main texts for details.
}
\label{fig:mcp}
\end{figure*}

We show the sensitivity reach of FLArE in the search for mCPs scattering off electrons in the detector in the left panel of \cref{fig:mcp}, together with previous constraints shown as gray-shaded regions and projections of other future experiments. In the plot, the red solid line corresponds to a single-scattered electron and the recoil energy cuts of $30~\mev \lesssim E_{R,\textrm{single}} \lesssim 1~\gev$. As can be seen, at the low $m_\chi$ end FLArE is able to compete with previous intensity frontier experiments such as LSND~\cite{Magill:2018tbb} and SLAC mQ~\cite{Prinz:1998ua}, while for the mass range between $100~\mev$ and $\mathcal{O}(100~\gev)$ it can go beyond the past bounds from ArgoNeuT~\cite{ArgoNeuT:2019ckq}, BEBC~\cite{Marocco:2020dqu}, CMS~\cite{CMS:2012xi,CMS:2013czn}, LEP~\cite{Davidson:2000hf}, and the milliQan prototype~\cite{milliQan:2021lne}.

When compared with future projections, although the FORMOSA detector~\cite{Foroughi-Abari:2020qar} to operate in the FPF is the leading proposal based on its projection, the FLArE search based on electron scatterings could provide an independent measurement method for a wide range of masses for mCP with charges $\mathcal{O}(10^{-3}\text{--}10^{-1})$. It can also have better sensitivity in the mass range above $\sim$ 4 GeV than proposed FerMINI~\cite{Kelly:2018brz} and SUBMET~\cite{Choi:2020mbk,Kim:2021eix} detectors. In the high-mass regime, FLArE is complementary to the sensitivity of milliQan~\cite{Haas:2014dda, Ball:2016zrp, Ball:2020dnx, milliQan:2021lne}, although it has a weaker sensitivity for $m_\chi\lesssim  (40\text{--}50)~\gev$. 

Importantly, while the other aforementioned proposed experiments will be targeting soft scintillation signals expected from mCPs traversing the detectors, the considered FLArE search is focused on larger energy deposits which can also be occasionally expected from energetic mCPs produced at the LHC. In order to highlight this even better, we also show in the plot with the red dashed curve the expected FLArE sensitivity assuming an increased lower recoil energy threshold of $300~\mev$, cf. \cref{eq:singlecuts}. In this case, the sensitivity reach becomes weaker, as expected for mCPs favoring soft electron recoils in the detector, cf. \cref{eq:scattering_xsec} for the scattering cross section. However, it can still constrain important and currently allowed regions in the mCP parameter space. 

Switching to multiple electron recoils, we find that the double-hit signal does not improve the sensitivity as the assumed detection threshold of $5~\mev$ is not low enough to identify smaller energy depositions characteristic for mCPs. This is in contrast to the scintillator-based detector of FORMOSA, which is able to detect even $\mathcal{O}({\rm eV})$ energy depositions. To similarly utilize the enhancement of event rate at low $E_R$ at FLArE, one needs a strategy that allows one to dial down the detection threshold, for example, a faint track signature. We leave this for future dedicated studies. 

In the right panel of \cref{fig:mcp}, we present a similar FLArE sensitivity due to single- and double-hit signals but assuming mCPs to be constituents of DM in the Universe. In this case, interesting effects on early-Universe cosmology and additional constraints from direct direction searches apply~\cite{Erickcek:2007jv, Rich:1987st}. We present the latter with light gray-shaded regions in the plot assuming the mCP abundance to be $0.4~\%$ of the total DM relic density to avoid strong cosmological constraints~\cite{Dubovsky:2003yn, Dolgov:2013una, Kovetz:2018zan}. The parameter space we consider corresponds to the ``strongly interacting dark matter'' region~\cite{Rich:1987st,Mahdawi:2018euy, Emken:2019tni, Foroughi-Abari:2020qar}, in the sense that the dark matter flux is attenuated through scattering in the atmosphere and crust. The mCP therefore loses its energy before getting to the terrestrial direct-detection experiments. Labeled as ``Direct Detection'' curve in the right panel of \cref{fig:mcp}, we show such a curve of critical cross-section. Still, a dedicated balloon-based experiment dedicated to search for such strongly interacting dark matter~\cite{Rich:1987st} (labeled as RRS), and a rocket-based experiment, X-ray Quantum Calorimeter (XQC)~\cite{McCammon:2002gb} provide constraints above this curve as they can detect DM particles above the atmosphere (RRS and XQC data are recast as bounds in~\cite{Mahdawi:2018euy}). FLArE will provide a strong probe for the region above this direct detection curve, independent of the assumptions of DM abundance as mCP.

In addition, we also highlight slices of the parameter space which were used to explain the aforementioned EDGES anomaly~\cite{Bowman:2018yin, Barkana:2018lgd, Berlin:2018bsc, Slatyer:2018aqg, Liu:2019knx} in \cref{fig:mcp}. We show two interesting bands of parameter space. One light green band corresponds to a minimal scenario which mCP makes up 0.4$\%$ of dark matter, without additional interactions to the rest of the cold DM. We also plot two scenarios in which 0.01$\%$ (10$^{-4}$) of dark matter is mCP, and the mCP DM is coupled to the rest of the cold DM through a light mediator to achieve additional cooling of the gas to explain the anomalous absorption spectrum observed by the EDGES collaboration. FLArE, with both single- and double-hit considerations, can study this parameter space close to the EDGES predictions.
 
\subsection{Dark states with electromagnetic form factors}
\label{sec:sensitivity_EMff}

\begin{figure*}[t]
\centering
\includegraphics[width=0.49\textwidth]{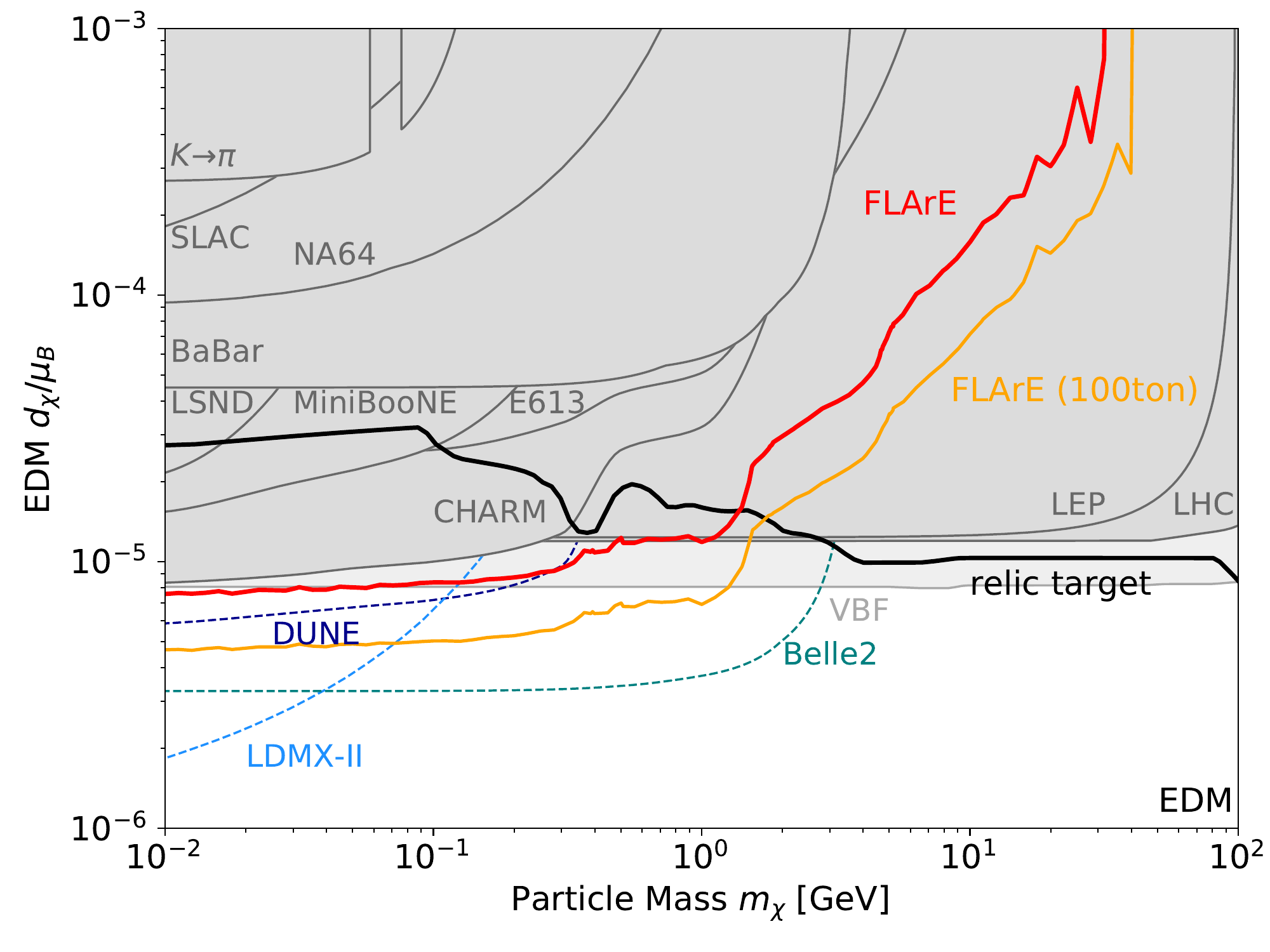}
\includegraphics[width=0.49\textwidth]{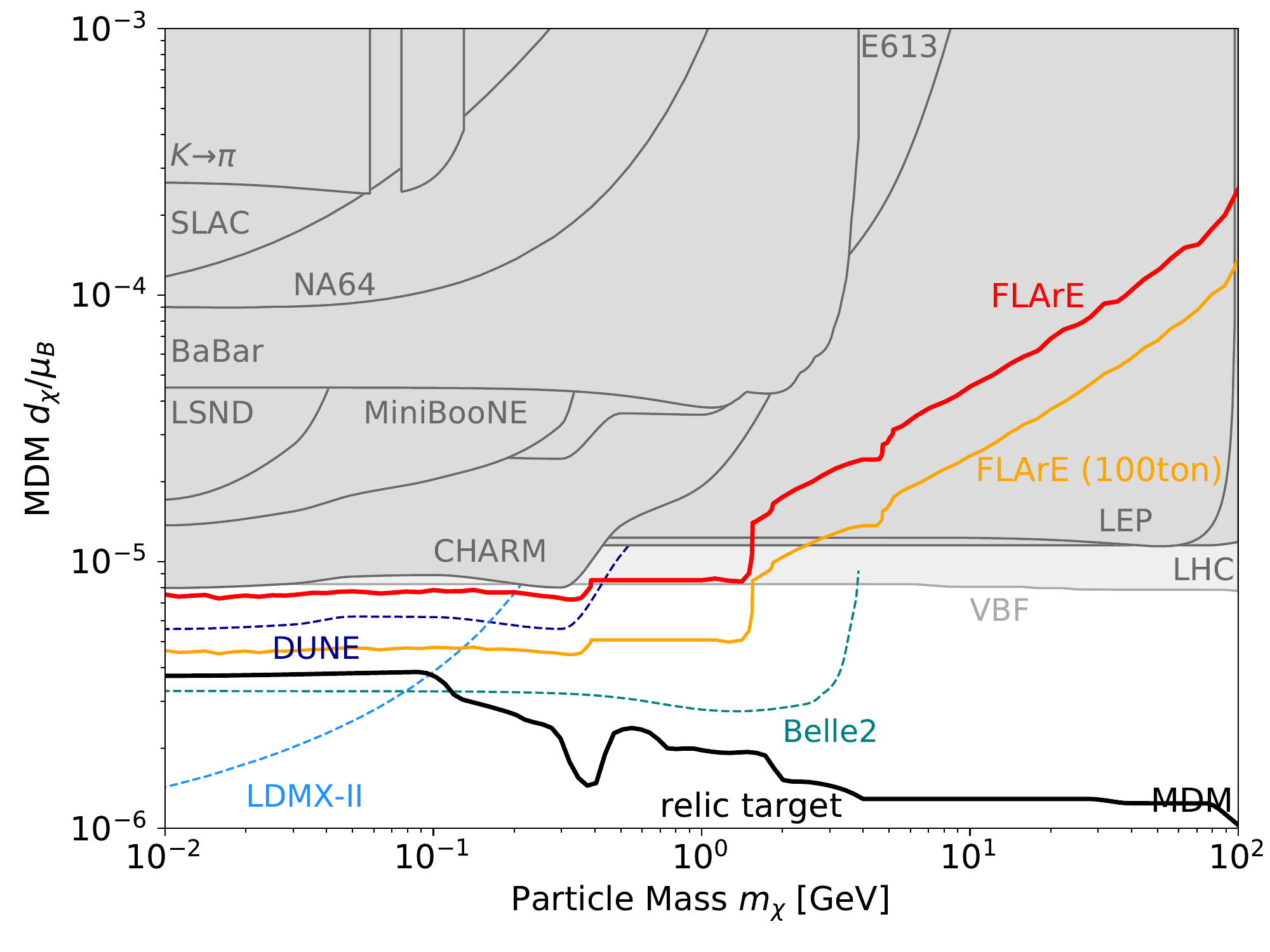}
\includegraphics[width=0.49\textwidth]{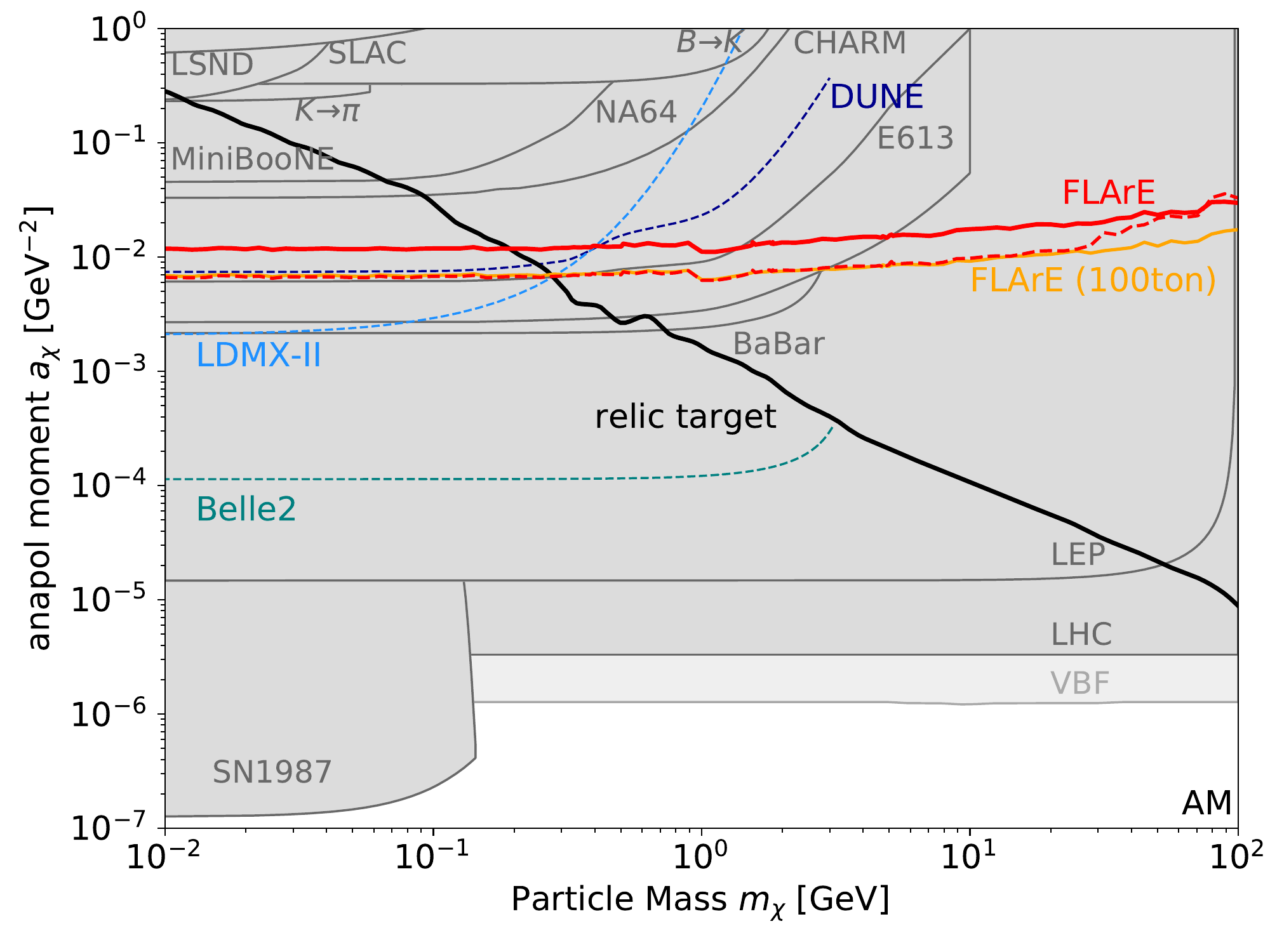}
\includegraphics[width=0.49\textwidth]{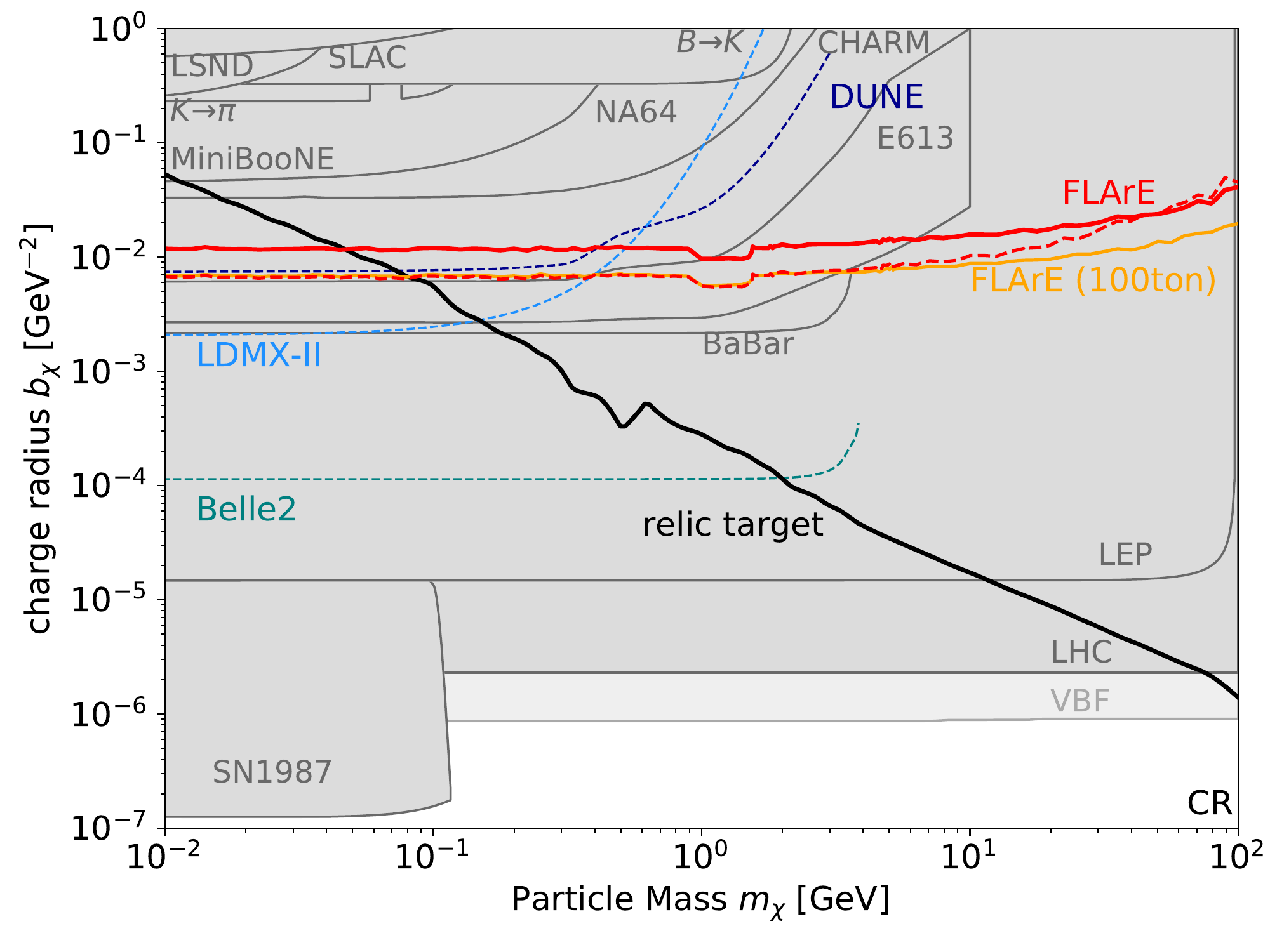}
\caption{Sensitivity of FLArE together with other accelerator/collider searches on $\chi$ with EM form factors. We can see that for mass-dimension $5$, FLArE can probe current unconstrained parameter space and be competitive with other proposed experiments. On the other hand, for mass-dimension $6$ searches focusing on electron recoil are in general superseded by LEP and LHC missing energy searches. We also show the bound from a search for missing-energy in vector boson fusion (VBF) processes; however, it needs to be taken with caveat that, in this regime, the EFT based on $F_{\mu\nu}$-only is not the appropriate description of the underlying interaction~\cite{Arina:2020mxo}.} 
\label{fig:sensitivity_EMDM}
\end{figure*}

For higher-dimensional EM form factor interactions, we focus on a single electron-recoil signal, since the scattering cross section does not favor low-energy deposition compared to that of mCP. The sensitivity reach of FLArE with 10-tonne (red solid curve) and 100-tonne (orange solid curve) detector for each EM form factor is shown in \cref{fig:sensitivity_EMDM}. In the top panels, we present the results for the mass-dimension $5$ operators, while the sensitivity reach for mass-dimension $6$ couplings is shown in the bottom panels. In each case, current bounds are shown with gray-shaded regions. Light dark species $\chi$ with $m_\chi\lesssim \mathcal{O}(10~\mev)$ are constrained dominantly by the past searches at the CHARM experiment~\cite{CHARM-II:1989nic,CHARM-II:1994dzw} in the mass-dimension $5$ case, and by supernovae bounds~\cite{Chu:2019rok} for the mass-dimension $6$ operators. Instead, heavier $\chi$s are constrained by mono-jet missing energy searches at the LHC~\cite{CMS:2017zts,Arina:2020mxo}. In the plots, we also present other bounds on the considered effective operators from the BaBar~\cite{BaBar:2001yhh}, E613~\cite{Ball:1980ojt}, LSND~\cite{LSND:1996jxj}, MiniBooNE~\cite{MiniBooNEDM:2018cxm}, NA64~\cite{NA64:2017vtt}, and SLAC beam-dump~\cite{Prinz:1998ua} experiments, as well as from the LEP L3 collaboration data~\cite{L3:2003yon,Fortin:2011hv} following Refs~\cite{Chu:2018qrm,Chu:2020ysb}. In addition, bounds from rare meson ($B$ and $K$) decay are included; see~\cite{Chu:2018qrm} and the references therein.  

We also show possibly more stringent constraints from di-jet searches at the LHC in which $\chi$s are produced via the vector boson fusion (VBF) process, cf. discussion in Ref.~\cite{Arina:2020mxo}. This production mode is enhanced in the presence of the $\chi$ couplings to only photons, although it also corresponds to the regime of this model in which unitarity might be violated. For this reason, we present this bound with a light gray color. We note that the relevant constraints become weaker than the aforementioned mono-jet bounds if the full hypercharge coupling is taken into account. In this case, however, further LEP bounds from invisible $Z$ boson decays would again render the relevant region in the parameter space excluded. 

In the case of mass-dimension $5$ operators shown on top, we find that FLArE can probe comparable EM form factor couplings compared to the past proton-beam experiments such as CHARM-II, and yield competitive sensitivity with the LHC search based on the VBF production. In order to probe unconstrained regions in the parameter space of these models, a $100$-tonne detector would be required. This will set new bounds for sub-GeV $\chi$, while for larger $\chi$ masses the expected sensitivity becomes weaker as the dark states can no longer be effectively produced in rare $J/\psi$ meson decays, cf. \cref{fig:prod}. In this case, missing energy searches at LEP and LHC become the most important. For reference, we also present the expected sensitivities of future proposed searches at the Belle-II, DUNE, and LDMX-II detectors, following Refs~\cite{Chu:2018qrm,Chu:2020ysb}.

Instead, for the mass-dimension $6$ operators, combination of SN1987A energy-loss bound~\cite{Chu:2019rok} and constraints from missing energy searches at colliders~\cite{Chu:2018qrm,Arina:2020mxo} cover most of the parameter space down to $a_\chi, b_\chi \sim 10^{-6}\,{\rm GeV}^{-2}$, while both FLArE and FLArE-100 only probe larger values of these couplings, $a_\chi, b_\chi \sim 10^{-2}\,{\rm GeV}^{-2}$. In general, missing-energy searches do not require produced dark states to have recoil signals, thus they, naively, win over searches based on the scattering signals due to the lack of an additional squared coupling suppression, assuming the same intensity or luminosity. 

We note, however, a relative difference between the mass-dimension $5$ and $6$ operators in the strength of these bounds compared to the FLArE sensitivity. This can be explained in a twofold way related to both the expected decreased FLArE $\chi$ detection prospects and increased sensitivity of missing-energy searches for $\chi$s coupled via the AM/CR couplings (dim-$6$) compared to lower-dimensional operators. As long as FLArE is concerned, this is first due to a suppressed $\chi$-production rate which, in the AM/CR case, is dominated by heavier $J/\psi$ mesons even for very forward-going $\chi$s, cf. right panels of \cref{fig:prod} for comparison (note different normalization factors used for different couplings). Moreover, the $E_R$-scaling of the differential scattering cross section for the mass-dimension $6$ operators is the same as that for neutrino-electron interactions. As a result, in the search based on low-energy electron recoils, a good fraction of the BSM-induced signal is cut together with background events. Second, the prospects for missing-energy searches for $\chi$s coupled via the mass-dimension $6$ operators are better than in the EDM/MDM case. This relies on the energy dependence of the $\chi$ production cross section for the process $e^- e^+ \rightarrow \chi \bar{\chi} \gamma$ which is increasingly proportional to the CM energy for higher-dimensional operators. These bounds will be further improved in the future HL-LHC searches~\cite{Arina:2020mxo}. 

We note that the FLArE detection prospects for $\chi$s coupled via the AM/CR operators could be improved by employing high-energy signal and nuclear scatterings, cf. Ref.~\cite{Batell:2021aja} for a similar discussion for the light DM search in FLArE. Importantly, for these types of interactions, increased target mass enhances the $\chi$ scattering cross section, cf. \cref{eq:scattering_xsec}. While signal and background differentiation is less trivial in this case, in order to reduce systematic uncertainties, new physics events could be searched for as an excess in neutral current (NC) over charged current (CC) neutrino-induced events (NC/CC). We leave a detailed analysis of this effect for future studies. We do not expect, however, this search strategy to improve FLArE sensitivity by additional several orders of magnitude in the coupling strength which is needed to probe unconstrained regions in the parameter space of these models. 

Last but not least, we note that $\chi$s coupled to photons via dimension $5$ and $6$ operators can also play the role of DM. In fact, they can be the thermal DM relic produced in the early Universe through the standard freeze-out mechanism; see solid black curves in~\cref{fig:sensitivity_EMDM} for the corresponding relic targets~\cite{Chu:2018qrm, Arina:2020mxo}. As can be seen, current bounds exclude the relevant regions of the parameter space of most of the considered scenarios for $m_\chi < \mathcal{O}(100\,{\rm GeV})$ with a notable exception found for the MDM model. In this case, the FLArE and FLArE-100 detectors could probe values of the coupling constant for sub-GeV $\chi$s for which a subdominant DM thermal relic density is predicted. In the case of the EDM model, the currently allowed region in the parameter space to be probed in FLArE/FLArE-100 corresponds to too small values of the annihilation cross section and, therefore, predicts too large thermal DM relic density. In order to reconcile such scenarios with the cosmology, one would have to either modify the cosmological evolution of the Universe after $\chi$s freeze out or introduce changes in the model by, e.g., adding intermediate metastable particles into which $\chi$s could preferentially annihilate to suppress their abundance. This would likely introduce further bounds but possibly also detection prospects for such a model in the FPF which can go beyond the simplest scenario we focus on in this study.

We also stress that $\chi$ DM for mass-dimension $5$ and $6$ operators can also be probed in direct detection (DD) experiments and indirect detection (ID) searches, cf. Ref.~\cite{Chu:2018qrm,Arina:2020mxo} for recent discussion. Although we do not show this in~\cref{fig:sensitivity_EMDM}, we note that these bounds can fully exclude the DM thermal freeze-out scenario of all the studied models for the mass range considered in this work. The only exception could be in the low-mass (sub-GeV) regime in which DD bounds become weaker, while ID constraints might be suppressed for a subdominant thermal DM relic density of $\chi$. To go beyond the aforementioned constraints, one can consider the freeze-in mechanism that the $\chi$ relic is sourced from feeble interactions in the early universe~\cite{Chang:2019xva}.

\section{Conclusion and Discussion}
\label{sec:discussion}

The testing of the coupling strength between the dark sector species and SM photons plays a particularly important role in searches for new physics. This is first due to their possible impact on astrophysical and cosmological probes of such scenarios, but also because such couplings might naturally arise in UV models, albeit they are typically suppressed, consistent with current experimental bounds. It is then essential to study new ways to systematically constrain such couplings.

In this study, we have discussed the discovery potential of such searches in the far-forward region of the LHC. Specifically, we have focused on the recently proposed Forward Physics Facility~\cite{MammenAbraham:2020hex, Anchordoqui:2021ghd, Feng:2022inv} and the FLArE experiment~\cite{Batell:2021blf, Feng:2022inv} in which stable dark species could be searched for via their scatterings off electrons in the detector. We have studied the detection prospects in the search for popular milli-charged particles (mCPs), but also assuming that dark sector particles are electrically neutral and coupled to the electromagnetic field strength tensor $F_{\mu\nu}$ only via higher-dimensional operators. In the latter case, we have analyzed the EM form factors induced by mass-dimension~$5$ electric (EDM) and magnetic (MDM) dipole moments, as well as on the anapole moment (AM) and charge radius (CR) at mass dimension~$6$.

We find good detection prospects in the search for single-scattered electrons produced in interactions of the mCPs. In this case, FLArE can probe currently allowed regions of the parameter space of this model for the dark species mass in between $10~\mev$ and a few tens of GeV. We have also analyzed the mCP detection prospects in FLArE based on two coincident and collinear soft hits induced by mCP scatterings off electrons. This search strategy remains particularly important for the mCP mass of order tens of GeV or so. The FLArE searches could independently contribute to strengthen such expected bounds from the search based on low ionization signals in the FORMOSA experiment~\cite{Foroughi-Abari:2020qar,Feng:2022inv}. In addition, we stress that FLArE capabilities to study the statistics of signal events with different electron recoil energies could provide additional information to better reconstruct the mCP parameters in the case of their discovery during the HL-LHC phase. Together, this places the proposed FPF among the best-suited facilities to search for mCPs in a wide range of masses and, simultaneously, using various experimental techniques.

It is important to mention that FPF searches for mCPs in both FLArE and FORMOSA could gain from the presence of both detectors placed along the beam axis in the FPF. In particular, this could help with better triggering the events and with vetoing muon-induced backgrounds therefore further strengthening the expected future bounds. It also remains interesting to study potential FLArE discovery prospects for mCPs via observations of dim tracks in the LArTPC detector, cf. Ref.~\cite{Harnik:2019zee} for a similar discussion. We leave such analyses for future studies.

While the detection prospects in the FPF based on scattering signatures are very promising in models predicting low-energy depositions in the detectors, scenarios preferring large momentum transfer in the interaction are more difficult to disentangle from neutrino-induced and other background sources. We illustrate this with the sensitivity reach of FLArE in the searches for EDM/MDM  dark sector particles. This remains comparable to the current bounds from proton beam-dump experiments, while for the $100$-tonne detector it can extend towards the future proposed searches in experimental facilities like Belle-II, DUNE, or LDMX-II. Even higher-dimensional operators leading to AM/CR interactions are more difficult to probe, with the expected FLArE reach not improving over the much stronger current bounds from the missing energy searches at LEP and LHC. In this case, an improved sensitivity reach could be obtained for more rich dark sector scenarios employing light new physics species and their couplings to the SM photons which could be studied via decay signatures in the FASER/FASER 2 experiments~\cite{FASER:2018ceo,FASER:2018bac} and thanks to the secondary production processes, cf. Refs~\cite{Jodlowski:2019ycu,Jodlowski:2020vhr}.

Last but not least, we stress that the results presented in the study have been obtained with the updated version of the \texttt{FORESEE} package~\cite{Kling:2021fwx}. On top of the previously implemented decay signatures, the package allows now for studying scattering signatures and contains further new physics models, as can be found in the public repository\footnote{\href{https://github.com/KlingFelix/FORESEE}{\faIcon{github} https://github.com/KlingFelix/FORESEE}}. Besides the LHC, the package can also be used to analyze the detection prospects of mCPs and other dark sector species in potential far-forward searches at future hadron colliders with $pp$ collisions at $27$ and $100~\tev$.

\vspace*{0.3cm}

\paragraph*{\bf Acknowledgments.}

We thank Andrew Cheek, Matthew Citron, Xiaoyong Chu, Milind Diwan, Jonathan Feng, Saeid Foroughi-Abari, Christopher Hill, Zhen Liu, Ornella Palamara, Josef Pradler, Adam Ritz, and David Stuart for useful discussions. We are grateful to the authors and maintainers of many open-source software packages, including
\texttt{CRMC}~\cite{CRMC}, 
\texttt{EPOS-LHC}~\cite{Pierog:2013ria},
\texttt{FORESEE}~\cite{Kling:2021fwx},
\texttt{MadGraph~5}~\cite{Alwall:2014hca},  \texttt{Pythia~8}~\cite{Sjostrand:2006za, Sjostrand:2014zea} and \texttt{scikit-hep}~\cite{Rodrigues:2019nct, Rodrigues:2020syo} 
including its \texttt{pylhe}~\cite{pylhe} and \texttt{Vector}~\cite{henry_schreiner_2022_5942083} libraries.
FK acknowledges support by the Deutsche Forschungsgemeinschaft under Germany's Excellence Strategy - EXC 2121 Quantum Universe - 390833306. JLK and YDT are supported by U.S. National Science Foundation (NSF) Theoretical Physics Program, Grant PHY-1915005. ST is supported by the grant ``AstroCeNT: Particle Astrophysics Science and Technology Centre'' carried out within the International Research Agendas programme of the Foundation for Polish Science financed by the European Union under the European Regional Development Fund. ST is supported in part by the Polish Ministry of Science and Higher Education through its scholarship for young and outstanding scientists (decision no 1190/E-78/STYP/14/2019). 
Part of this manuscript has been authored by Fermi Research Alliance, LLC under Contract No. DE-AC02-07CH11359 with the U.S. Department of Energy, Office of Science, Office of High Energy Physics. YDT thank the Kavli Institute for Theoretical Physics at the University of California, Santa Barbara, for its excellent program "Neutrinos as a Portal to New Physics and Astrophysics," supported in part by the National Science Foundation under Grant No. NSF PHY-1748958. This work was partially performed at the Aspen Center for Physics, which is supported by National Science Foundation grant PHY-1607611.

\appendix

\section{Pseudoscalar and vector meson decay}
\label{app:meson_decay}

In this appendix, we list relevant formulas the production of the dark sector state $\chi$ in meson decays. For pseudoscalar decay, we use the differential branching ratio which can be expressed as 
\begin{widetext}
\be
\label{eq:br_scalar}
\text{mCP:}\quad
   &\frac{d{\rm BR}_{P \rightarrow \gamma\chi\bar{\chi}}}{dq^2 d\cos\theta} = {\rm BR}_{P\rightarrow \gamma \gamma}  \!\times\!\! \left[ \frac{\alpha\epsilon^2}{4\pi q^2} \left( 1 \!-\! \frac{q^2}{M^2} \right)^3 \!\!\sqrt{1\!-\!\frac{4m_\chi^2}{q^2}} \left( 2 \!-\! \left(1 \!-\! \frac{4m_\chi^2}{q^2} \right) \sin^2\theta \right) \!\right]  \\
\text{MDM:}\quad     
	&  \frac{d{\rm BR}_{P \rightarrow \gamma\chi\bar{\chi}}}{dq^2 d\cos\theta} = {\rm BR}_{P\rightarrow \gamma \gamma}  \!\times\!\!  \left[ \frac{\mu_\chi^2}{16\pi^2 } \!\left( \!1 \!-\! \frac{q^2}{M^2}\! \right)^3 \!\!\!\sqrt{1\!-\!\frac{4m_\chi^2}{q^2}} \left(\! \frac{8m_\chi^2}{q^2} \!-\! \left[1 \!-\! \frac{4m_\chi^2}{q^2} \right] \sin^2\theta \!\right)\! \!\right]  \\
\text{EDM:}\quad     
	&  \frac{d{\rm BR}_{P \rightarrow \gamma\chi\bar{\chi}}}{dq^2 d\cos\theta} = {\rm BR}_{P\rightarrow \gamma \gamma}  \!\times\!\! \left[ \frac{d_\chi^2}{16\pi^2 } \left( 1 \!-\! \frac{q^2}{M^2} \right)^3 \!\!\sqrt{1\!-\!\frac{4m_\chi^2}{q^2}}  \left(1 \!-\! \frac{4m_\chi^2}{q^2} \right) \sin^2\theta \right]  \\
\text{AM:}\quad     
	&  \frac{d{\rm BR}_{P \rightarrow \gamma\chi\bar{\chi}}}{dq^2 d\cos\theta} = {\rm BR}_{P\rightarrow \gamma \gamma}  \!\times\!\! \left[ \frac{a_\chi^2 q^2}{16\pi^2 } \left( 1 \!-\! \frac{q^2}{M^2} \right)^3 \!\!\sqrt{1\!-\!\frac{4m_\chi^2}{q^2}}  \left(1 \!-\! \frac{4m_\chi^2}{q^2} \right)(2 \!-\! \sin^2\theta) \! \right] \\
\text{CR:}\quad     
	&  \frac{d{\rm BR}_{P \rightarrow \gamma\chi\bar{\chi}}}{dq^2 d\cos\theta} = {\rm BR}_{P\rightarrow \gamma \gamma}  \!\times\!\! \left[ \frac{b_\chi^2 q^2}{16\pi^2 } \left( 1 \!-\! \frac{q^2}{M^2} \right)^3 \!\!\sqrt{1\!-\!\frac{4m_\chi^2}{q^2}}   \left( 2 \!-\! \left[1 \!-\! \frac{4m_\chi^2}{q^2} \right] \sin^2\theta \right) \!\right]	
\ee
\end{widetext}
where $M$ is the mass of pseudoscalar meson, $q^2$ is the invariant mass square of $\chi$-pair, $\theta$ is the angle between $p_\chi$ in $\chi$-pair rest frame and $p_\chi+p_{\bar{\chi}}$ in meson rest frame, and ${\rm BR}_{P\rightarrow \gamma \gamma}$ is the branching ratio of pseudoscalar meson decaying into two photons.
We adopt the values of ${\rm BR}_{P\rightarrow \gamma \gamma}$ reported in PDG~\cite{ParticleDataGroup:2020ssz}.
For vector meson decay, we implement the branching ratio into $\chi$-pair for each millicharge and EM form factor rescaled from the branching ratio into electron-positron pair ${\rm BR}_{V\rightarrow ee}$, which reads
\be
\!\!\text{mCP:}\     
	&\frac{{\rm BR}_{V \rightarrow \chi\bar{\chi}}}{{\rm BR}_{V \rightarrow ee}} \!=\! \epsilon^2 \frac{M^2 \!+\!2 m_\chi^2}{M^2 \!+\!2 m_e^2} \sqrt{\frac{M^2 \!-\!4 m_\chi^2}{M^2 \!-\!4 m_e^2}} ,\!\!\\
\!\!\text{MDM:}\     
	&\frac{{\rm BR}_{V \rightarrow \chi\bar{\chi}}}{{\rm BR}_{V \rightarrow ee}} \!=\! \dfrac{\mu_\chi^2 M^2}{8\pi\alpha} \frac{M^2 \!+\!8 m_\chi^2}{M^2 \!+\!2 m_e^2} \sqrt{\frac{M^2 \!-\!4 m_\chi^2}{M^2 \!-\!4 m_e^2}} ,\!\! \\
\!\!\text{EDM:}\     
	&\frac{{\rm BR}_{V \rightarrow \chi\bar{\chi}}}{{\rm BR}_{V \rightarrow ee}} \!=\! \dfrac{d_\chi^2 M^2}{8\pi\alpha} \frac{M^2 \!-\!4 m_\chi^2}{M^2 \!+\!2 m_e^2} \sqrt{\frac{M^2 \!-\!4 m_\chi^2}{M^2 \!-\!4 m_e^2}} ,\!\! \\
\!\!\text{AM:}\     
	&\frac{{\rm BR}_{V \rightarrow \chi\bar{\chi}}}{{\rm BR}_{V \rightarrow ee}} \!=\! \dfrac{a_\chi^2 M^4}{4\pi\alpha} \frac{M^2 \!-\!4 m_\chi^2}{M^2 \!+\!2 m_e^2} \sqrt{\frac{M^2 \!-\!4 m_\chi^2}{M^2 \!-\!4 m_e^2}} ,\!\! \\
\!\!\text{CR:}\     
	&\frac{{\rm BR}_{V \rightarrow \chi\bar{\chi}}}{{\rm BR}_{V \rightarrow ee}} \!=\! \dfrac{b_\chi^2 M^4}{4\pi\alpha} \frac{M^2 \!+\!2 m_\chi^2}{M^2 \!+\!2 m_e^2} \sqrt{\frac{M^2 \!-\!4 m_\chi^2}{M^2 \!-\!4 m_e^2}}  ,\!\!
    \label{eq:br_vector}
\ee
with $M$ being the mass of vector meson and the benchmark values of ${\rm BR}_{V\rightarrow ee}$ taken from PDG~\cite{ParticleDataGroup:2020ssz}.

\section{DM-electron scattering cross section}

The differential scattering cross section for $\chi e\to \chi e$ reads
\be
\label{eq:differential_xsec}
	\frac{d\sigma}{dE_R} 
		=  \alpha \frac{  g_E(E_R) + g_M(E_R) }{(E_{\chi }^2-m_{\chi}^2) (2 m_e+E_R) }\,,
\ee
with the functions $g_E(E_R)$ and $g_M(E_R)$ for all EM interactions can be found in~\cite{Chu:2018qrm}.
Given that $\chi$ particles are highly-boosted for considered mass range, we can simplify \cref{eq:differential_xsec} assuming that $E_\chi \gg m_e, m_\chi, E_R$ and $E_R \gg m_e$. 
For each EM form factor, the differential cross section that we adopt in \texttt{FORESEE} can be expressed as 
\be
\label{eq:scattering_xsec}
 \text{mCP:} \       
  &\frac{d\sigma}{dE_R} = \frac{4\pi\alpha^2 \epsilon^2}{E_{\chi }^2  } \times \left(
                          \frac{E_{\chi }^2}{2 m_e E_R^2 } - \frac{m_{\chi }^2}{4 m_e^2 E_R}  \right),\\
 \text{MDM:}  \  
 &\frac{d\sigma}{dE_R} = \frac{\alpha \mu_\chi^2}{E_{\chi }^2 } \times \left(
                         \frac{ m_\chi^2}{2 m_e} +  \frac{E_\chi^2}{E_R }  \right), \\
 \text{EDM:}   \  
 &\frac{d\sigma}{dE_R} = \frac{\alpha d_\chi^2}{E_{\chi }^2  } \times \left(-
                         \frac{ m_\chi^2}{2 m_e} +  \frac{E_\chi^2}{E_R }  \right) ,\\
 \text{AM:} \     
 &\frac{d\sigma}{dE_R} = \frac{\alpha a_\chi^2 }{E_{\chi }^2  } \times \left(
                          2 m_e  E_\chi^2 + E_R m_{\chi }^2   \right), \\
 \text{CR:}  \   
 &\frac{d\sigma}{dE_R} =  \frac{\alpha b_\chi^2 }{E_{\chi }^2 } \times \left(
                          2 m_e  E_{\chi }^2 - E_R  m_{\chi }^2  \right).
\ee
Assuming electrons are at rest in the lab frame, the maximal recoil energy $E_R^{\rm max}$ given $E_\chi$ reads
\be
 E_R^{\rm max}  = \dfrac{2m_e (E_\chi^2 - m_\chi^2)}{m_e (2E_\chi + m_e) + m_\chi^2} \approx \frac{2 m_e E_\chi^2}{2 m_e E_\chi + m_\chi^2}\,.
\ee

\bibliography{references}

\providecommand{\href}[2]{#2}\begingroup\raggedright\begin{thebibliography}{100}

\bibitem{Pati:1973uk}
J.~C. Pati and A.~Salam, ``{Unified Lepton-Hadron Symmetry and a Gauge Theory
  of the Basic Interactions},''
  \href{http://dx.doi.org/10.1103/PhysRevD.8.1240}{{\em Phys. Rev. D} {\bf 8}
  (1973)  1240--1251}.

\bibitem{Georgi:1974my}
H.~Georgi, ``{The State of the Art\textemdash{}Gauge Theories},''
  \href{http://dx.doi.org/10.1063/1.2947450}{{\em AIP Conf. Proc.} {\bf 23}
  (1975)  575--582}.

\bibitem{Wen:1985qj}
X.-G. Wen and E.~Witten, ``{Electric and Magnetic Charges in Superstring
  Models},'' \href{http://dx.doi.org/10.1016/0550-3213(85)90592-9}{{\em Nucl.
  Phys. B} {\bf 261} (1985)  651--677}.

\bibitem{Shiu:2013wxa}
G.~Shiu, P.~Soler, and F.~Ye, ``{Milli-Charged Dark Matter in Quantum Gravity
  and String Theory},''
  \href{http://dx.doi.org/10.1103/PhysRevLett.110.241304}{{\em Phys. Rev.
  Lett.} {\bf 110} (2013) no.~24, 241304},
  \href{http://arxiv.org/abs/1302.5471}{{\tt arXiv:1302.5471 [hep-th]}}.

\bibitem{Holdom:1985ag}
B.~Holdom, ``{Two U(1)'s and Epsilon Charge Shifts},''
  \href{http://dx.doi.org/10.1016/0370-2693(86)91377-8}{{\em Phys. Lett. B}
  {\bf 166} (1986)  196--198}.

\bibitem{Davidson:2000hf}
S.~Davidson, S.~Hannestad, and G.~Raffelt, ``{Updated bounds on millicharged
  particles},'' \href{http://dx.doi.org/10.1088/1126-6708/2000/05/003}{{\em
  JHEP} {\bf 05} (2000)  003}, \href{http://arxiv.org/abs/hep-ph/0001179}{{\tt
  arXiv:hep-ph/0001179}}.

\bibitem{Pospelov:2007mp}
M.~Pospelov, A.~Ritz, and M.~B. Voloshin, ``{Secluded WIMP Dark Matter},''
  \href{http://dx.doi.org/10.1016/j.physletb.2008.02.052}{{\em Phys. Lett. B}
  {\bf 662} (2008)  53--61}, \href{http://arxiv.org/abs/0711.4866}{{\tt
  arXiv:0711.4866 [hep-ph]}}.

\bibitem{Vogel:2013raa}
H.~Vogel and J.~Redondo, ``{Dark Radiation constraints on minicharged particles
  in models with a hidden photon},''
  \href{http://dx.doi.org/10.1088/1475-7516/2014/02/029}{{\em JCAP} {\bf 02}
  (2014)  029}, \href{http://arxiv.org/abs/1311.2600}{{\tt arXiv:1311.2600
  [hep-ph]}}.

\bibitem{Rich:1987st}
J.~Rich, R.~Rocchia, and M.~Spiro, ``{A Search for Strongly Interacting Dark
  Matter},'' \href{http://dx.doi.org/10.1016/0370-2693(87)90788-X}{{\em Phys.
  Lett.} {\bf B194} (1987)  173}.
[,221(1987)].

\bibitem{Mahdawi:2018euy}
M.~S. Mahdawi and G.~R. Farrar, ``{Constraints on Dark Matter with a moderately
  large and velocity-dependent DM-nucleon cross-section},''
  \href{http://dx.doi.org/10.1088/1475-7516/2018/10/007}{{\em JCAP} {\bf 1810}
  (2018) no.~10, 007},
\href{http://arxiv.org/abs/1804.03073}{{\tt arXiv:1804.03073 [hep-ph]}}.

\bibitem{Emken:2019tni}
T.~Emken, R.~Essig, C.~Kouvaris, and M.~Sholapurkar, ``{Direct Detection of
  Strongly Interacting Sub-GeV Dark Matter via Electron Recoils},''
  \href{http://dx.doi.org/10.1088/1475-7516/2019/09/070}{{\em JCAP} {\bf 1909}
  (2019) no.~09, 070},
\href{http://arxiv.org/abs/1905.06348}{{\tt arXiv:1905.06348 [hep-ph]}}.

\bibitem{Foroughi-Abari:2020qar}
S.~Foroughi-Abari, F.~Kling, and Y.-D. Tsai, ``{Looking forward to millicharged
  dark sectors at the LHC},''
  \href{http://dx.doi.org/10.1103/PhysRevD.104.035014}{{\em Phys. Rev. D} {\bf
  104} (2021) no.~3, 035014}, \href{http://arxiv.org/abs/2010.07941}{{\tt
  arXiv:2010.07941 [hep-ph]}}.

\bibitem{Bowman:2018yin}
J.~D. Bowman, A.~E.~E. Rogers, R.~A. Monsalve, T.~J. Mozdzen, and N.~Mahesh,
  ``{An absorption profile centred at 78 megahertz in the sky-averaged
  spectrum},'' \href{http://dx.doi.org/10.1038/nature25792}{{\em Nature} {\bf
  555} (2018) no.~7694, 67--70}, \href{http://arxiv.org/abs/1810.05912}{{\tt
  arXiv:1810.05912 [astro-ph.CO]}}.

\bibitem{Munoz:2018pzp}
J.~B. Mu\~noz and A.~Loeb, ``{A small amount of mini-charged dark matter could
  cool the baryons in the early Universe},''
  \href{http://dx.doi.org/10.1038/s41586-018-0151-x}{{\em Nature} {\bf 557}
  (2018) no.~7707, 684}, \href{http://arxiv.org/abs/1802.10094}{{\tt
  arXiv:1802.10094 [astro-ph.CO]}}.

\bibitem{Fialkov:2018xre}
A.~Fialkov, R.~Barkana, and A.~Cohen, ``{Constraining Baryon--Dark Matter
  Scattering with the Cosmic Dawn 21-cm Signal},''
  \href{http://dx.doi.org/10.1103/PhysRevLett.121.011101}{{\em Phys. Rev.
  Lett.} {\bf 121} (2018)  011101}, \href{http://arxiv.org/abs/1802.10577}{{\tt
  arXiv:1802.10577 [astro-ph.CO]}}.

\bibitem{Berlin:2018sjs}
A.~Berlin, D.~Hooper, G.~Krnjaic, and S.~D. McDermott, ``{Severely Constraining
  Dark Matter Interpretations of the 21-cm Anomaly},''
  \href{http://dx.doi.org/10.1103/PhysRevLett.121.011102}{{\em Phys. Rev.
  Lett.} {\bf 121} (2018) no.~1, 011102},
  \href{http://arxiv.org/abs/1803.02804}{{\tt arXiv:1803.02804 [hep-ph]}}.

\bibitem{Barkana:2018qrx}
R.~Barkana, N.~J. Outmezguine, D.~Redigolo, and T.~Volansky, ``{Strong
  constraints on light dark matter interpretation of the EDGES signal},''
  \href{http://dx.doi.org/10.1103/PhysRevD.98.103005}{{\em Phys. Rev. D} {\bf
  98} (2018) no.~10, 103005}, \href{http://arxiv.org/abs/1803.03091}{{\tt
  arXiv:1803.03091 [hep-ph]}}.

\bibitem{Barkana:2018lgd}
R.~Barkana, ``{Possible interaction between baryons and dark-matter particles
  revealed by the first stars},''
  \href{http://dx.doi.org/10.1038/nature25791}{{\em Nature} {\bf 555} (2018)
  no.~7694, 71--74}, \href{http://arxiv.org/abs/1803.06698}{{\tt
  arXiv:1803.06698 [astro-ph.CO]}}.

\bibitem{Slatyer:2018aqg}
T.~R. Slatyer and C.-L. Wu, ``{Early-Universe constraints on dark matter-baryon
  scattering and their implications for a global 21 cm signal},''
  \href{http://dx.doi.org/10.1103/PhysRevD.98.023013}{{\em Phys. Rev. D} {\bf
  98} (2018) no.~2, 023013}, \href{http://arxiv.org/abs/1803.09734}{{\tt
  arXiv:1803.09734 [astro-ph.CO]}}.

\bibitem{Boddy:2018wzy}
K.~K. Boddy, V.~Gluscevic, V.~Poulin, E.~D. Kovetz, M.~Kamionkowski, and
  R.~Barkana, ``{Critical assessment of CMB limits on dark matter-baryon
  scattering: New treatment of the relative bulk velocity},''
  \href{http://dx.doi.org/10.1103/PhysRevD.98.123506}{{\em Phys. Rev. D} {\bf
  98} (2018) no.~12, 123506}, \href{http://arxiv.org/abs/1808.00001}{{\tt
  arXiv:1808.00001 [astro-ph.CO]}}.

\bibitem{Kovetz:2018zan}
E.~D. Kovetz, V.~Poulin, V.~Gluscevic, K.~K. Boddy, R.~Barkana, and
  M.~Kamionkowski, ``{Tighter limits on dark matter explanations of the
  anomalous EDGES 21 cm signal},''
  \href{http://dx.doi.org/10.1103/PhysRevD.98.103529}{{\em Phys. Rev. D} {\bf
  98} (2018) no.~10, 103529}, \href{http://arxiv.org/abs/1807.11482}{{\tt
  arXiv:1807.11482 [astro-ph.CO]}}.

\bibitem{Creque-Sarbinowski:2019mcm}
C.~Creque-Sarbinowski, L.~Ji, E.~D. Kovetz, and M.~Kamionkowski, ``{Direct
  millicharged dark matter cannot explain the EDGES signal},''
  \href{http://dx.doi.org/10.1103/PhysRevD.100.023528}{{\em Phys. Rev. D} {\bf
  100} (2019) no.~2, 023528}, \href{http://arxiv.org/abs/1903.09154}{{\tt
  arXiv:1903.09154 [astro-ph.CO]}}.

\bibitem{Liu:2019knx}
H.~Liu, N.~J. Outmezguine, D.~Redigolo, and T.~Volansky, ``{Reviving
  Millicharged Dark Matter for 21-cm Cosmology},''
  \href{http://dx.doi.org/10.1103/PhysRevD.100.123011}{{\em Phys. Rev. D} {\bf
  100} (2019) no.~12, 123011}, \href{http://arxiv.org/abs/1908.06986}{{\tt
  arXiv:1908.06986 [hep-ph]}}.

\bibitem{CMS:2012xi}
{\bf CMS} Collaboration, S.~Chatrchyan {\em et al.}, ``{Search for Fractionally
  Charged Particles in $pp$ Collisions at $\sqrt{s}=7$ TeV},''
  \href{http://dx.doi.org/10.1103/PhysRevD.87.092008}{{\em Phys. Rev. D} {\bf
  87} (2013) no.~9, 092008}, \href{http://arxiv.org/abs/1210.2311}{{\tt
  arXiv:1210.2311 [hep-ex]}}.

\bibitem{CMS:2013czn}
{\bf CMS} Collaboration, S.~Chatrchyan {\em et al.}, ``{Searches for Long-Lived
  Charged Particles in $pp$ Collisions at $\sqrt{s}$=7 and 8 TeV},''
  \href{http://dx.doi.org/10.1007/JHEP07(2013)122}{{\em JHEP} {\bf 07} (2013)
  122}, \href{http://arxiv.org/abs/1305.0491}{{\tt arXiv:1305.0491 [hep-ex]}}.

\bibitem{Jaeckel:2012yz}
J.~Jaeckel, M.~Jankowiak, and M.~Spannowsky, ``{LHC probes the hidden
  sector},'' \href{http://dx.doi.org/10.1016/j.dark.2013.06.001}{{\em Phys.
  Dark Univ.} {\bf 2} (2013)  111--117},
  \href{http://arxiv.org/abs/1212.3620}{{\tt arXiv:1212.3620 [hep-ph]}}.

\bibitem{Essig:2013lka}
R.~Essig {\em et al.}, ``{Working Group Report: New Light Weakly Coupled
  Particles},'' in {\em {Community Summer Study 2013}: {Snowmass on the
  Mississippi}}.
\newblock 10, 2013.
\newblock \href{http://arxiv.org/abs/1311.0029}{{\tt arXiv:1311.0029
  [hep-ph]}}.

\bibitem{Haas:2014dda}
A.~Haas, C.~S. Hill, E.~Izaguirre, and I.~Yavin, ``{Looking for milli-charged
  particles with a new experiment at the LHC},''
  \href{http://dx.doi.org/10.1016/j.physletb.2015.04.062}{{\em Phys. Lett. B}
  {\bf 746} (2015)  117--120}, \href{http://arxiv.org/abs/1410.6816}{{\tt
  arXiv:1410.6816 [hep-ph]}}.

\bibitem{Ball:2016zrp}
A.~Ball {\em et al.}, ``{A Letter of Intent to Install a milli-charged Particle
  Detector at LHC P5},'' \href{http://arxiv.org/abs/1607.04669}{{\tt
  arXiv:1607.04669 [physics.ins-det]}}.

\bibitem{milliQan:2021lne}
{\bf milliQan} Collaboration, A.~Ball {\em et al.}, ``{Sensitivity to
  millicharged particles in future proton-proton collisions at the LHC with the
  milliQan detector},''
  \href{http://dx.doi.org/10.1103/PhysRevD.104.032002}{{\em Phys. Rev. D} {\bf
  104} (2021) no.~3, 032002}, \href{http://arxiv.org/abs/2104.07151}{{\tt
  arXiv:2104.07151 [hep-ex]}}.

\bibitem{Ball:2020dnx}
A.~Ball {\em et al.}, ``{Search for millicharged particles in proton-proton
  collisions at $\sqrt{s} = 13$ TeV},''
  \href{http://dx.doi.org/10.1103/PhysRevD.102.032002}{{\em Phys. Rev. D} {\bf
  102} (2020) no.~3, 032002}, \href{http://arxiv.org/abs/2005.06518}{{\tt
  arXiv:2005.06518 [hep-ex]}}.

\bibitem{Liu:2018jdi}
Z.~Liu and Y.~Zhang, ``{Probing millicharge at BESIII via monophoton
  searches},'' \href{http://dx.doi.org/10.1103/PhysRevD.99.015004}{{\em Phys.
  Rev. D} {\bf 99} (2019) no.~1, 015004},
  \href{http://arxiv.org/abs/1808.00983}{{\tt arXiv:1808.00983 [hep-ph]}}.

\bibitem{Prinz:1998ua}
A.~A. Prinz {\em et al.}, ``{Search for millicharged particles at SLAC},''
  \href{http://dx.doi.org/10.1103/PhysRevLett.81.1175}{{\em Phys. Rev. Lett.}
  {\bf 81} (1998)  1175--1178}, \href{http://arxiv.org/abs/hep-ex/9804008}{{\tt
  arXiv:hep-ex/9804008}}.

\bibitem{Berlin:2018bsc}
A.~Berlin, N.~Blinov, G.~Krnjaic, P.~Schuster, and N.~Toro, ``{Dark Matter,
  Millicharges, Axion and Scalar Particles, Gauge Bosons, and Other New Physics
  with LDMX},'' \href{http://dx.doi.org/10.1103/PhysRevD.99.075001}{{\em Phys.
  Rev. D} {\bf 99} (2019) no.~7, 075001},
  \href{http://arxiv.org/abs/1807.01730}{{\tt arXiv:1807.01730 [hep-ph]}}.

\bibitem{Gninenko:2018ter}
S.~N. Gninenko, D.~V. Kirpichnikov, and N.~V. Krasnikov, ``{Probing
  millicharged particles with NA64 experiment at CERN},''
  \href{http://dx.doi.org/10.1103/PhysRevD.100.035003}{{\em Phys. Rev. D} {\bf
  100} (2019) no.~3, 035003}, \href{http://arxiv.org/abs/1810.06856}{{\tt
  arXiv:1810.06856 [hep-ph]}}.

\bibitem{Arefyeva:2022eba}
N.~Arefyeva, S.~Gninenko, D.~Gorbunov, and D.~Kirpichnikov, ``{Passage of
  millicharged particles in the electron beam-dump: refining constraints from
  SLACmQ and estimating sensitivity of NA64e},''
  \href{http://arxiv.org/abs/2204.03984}{{\tt arXiv:2204.03984 [hep-ph]}}.

\bibitem{Magill:2018tbb}
G.~Magill, R.~Plestid, M.~Pospelov, and Y.-D. Tsai, ``{Millicharged particles
  in neutrino experiments},''
  \href{http://dx.doi.org/10.1103/PhysRevLett.122.071801}{{\em Phys. Rev.
  Lett.} {\bf 122} (2019) no.~7, 071801},
  \href{http://arxiv.org/abs/1806.03310}{{\tt arXiv:1806.03310 [hep-ph]}}.

\bibitem{Kelly:2018brz}
K.~J. Kelly and Y.-D. Tsai, ``{Proton fixed-target scintillation experiment to
  search for millicharged dark matter},''
  \href{http://dx.doi.org/10.1103/PhysRevD.100.015043}{{\em Phys. Rev. D} {\bf
  100} (2019) no.~1, 015043}, \href{http://arxiv.org/abs/1812.03998}{{\tt
  arXiv:1812.03998 [hep-ph]}}.

\bibitem{Beacham:2019nyx}
J.~Beacham {\em et al.}, ``{Physics Beyond Colliders at CERN: Beyond the
  Standard Model Working Group Report},''
  \href{http://dx.doi.org/10.1088/1361-6471/ab4cd2}{{\em J. Phys. G} {\bf 47}
  (2020) no.~1, 010501}, \href{http://arxiv.org/abs/1901.09966}{{\tt
  arXiv:1901.09966 [hep-ex]}}.

\bibitem{Harnik:2019zee}
R.~Harnik, Z.~Liu, and O.~Palamara, ``{Millicharged Particles in Liquid Argon
  Neutrino Experiments},''
  \href{http://dx.doi.org/10.1007/JHEP07(2019)170}{{\em JHEP} {\bf 07} (2019)
  170}, \href{http://arxiv.org/abs/1902.03246}{{\tt arXiv:1902.03246
  [hep-ph]}}.

\bibitem{ArgoNeuT:2019ckq}
{\bf ArgoNeuT} Collaboration, R.~Acciarri {\em et al.}, ``{Improved Limits on
  Millicharged Particles Using the ArgoNeuT Experiment at Fermilab},''
  \href{http://dx.doi.org/10.1103/PhysRevLett.124.131801}{{\em Phys. Rev.
  Lett.} {\bf 124} (2020) no.~13, 131801},
  \href{http://arxiv.org/abs/1911.07996}{{\tt arXiv:1911.07996 [hep-ex]}}.

\bibitem{Marocco:2020dqu}
G.~Marocco and S.~Sarkar, ``{Blast from the past: Constraints on the dark
  sector from the BEBC WA66 beam dump experiment},''
  \href{http://dx.doi.org/10.21468/SciPostPhys.10.2.043}{{\em SciPost Phys.}
  {\bf 10} (2021) no.~2, 043}, \href{http://arxiv.org/abs/2011.08153}{{\tt
  arXiv:2011.08153 [hep-ph]}}.

\bibitem{Kavanagh:2018xeh}
B.~J. Kavanagh, P.~Panci, and R.~Ziegler, ``{Faint Light from Dark Matter:
  Classifying and Constraining Dark Matter-Photon Effective Operators},''
  \href{http://dx.doi.org/10.1007/JHEP04(2019)089}{{\em JHEP} {\bf 04} (2019)
  089}, \href{http://arxiv.org/abs/1810.00033}{{\tt arXiv:1810.00033
  [hep-ph]}}.

\bibitem{Magill:2018jla}
G.~Magill, R.~Plestid, M.~Pospelov, and Y.-D. Tsai, ``{Dipole Portal to Heavy
  Neutral Leptons},'' \href{http://dx.doi.org/10.1103/PhysRevD.98.115015}{{\em
  Phys. Rev. D} {\bf 98} (2018) no.~11, 115015},
  \href{http://arxiv.org/abs/1803.03262}{{\tt arXiv:1803.03262 [hep-ph]}}.

\bibitem{Shoemaker:2020kji}
I.~M. Shoemaker, Y.-D. Tsai, and J.~Wyenberg, ``{Active-to-sterile neutrino
  dipole portal and the XENON1T excess},''
  \href{http://dx.doi.org/10.1103/PhysRevD.104.115026}{{\em Phys. Rev. D} {\bf
  104} (2021) no.~11, 115026}, \href{http://arxiv.org/abs/2007.05513}{{\tt
  arXiv:2007.05513 [hep-ph]}}.

\bibitem{Brdar:2020quo}
V.~Brdar, A.~Greljo, J.~Kopp, and T.~Opferkuch, ``{The Neutrino Magnetic Moment
  Portal: Cosmology, Astrophysics, and Direct Detection},''
  \href{http://dx.doi.org/10.1088/1475-7516/2021/01/039}{{\em JCAP} {\bf 01}
  (2021)  039}, \href{http://arxiv.org/abs/2007.15563}{{\tt arXiv:2007.15563
  [hep-ph]}}.

\bibitem{Dasgupta:2021fpn}
A.~Dasgupta, S.~K. Kang, and J.~E. Kim, ``{Probing neutrino dipole portal at
  COHERENT experiment},'' \href{http://dx.doi.org/10.1007/JHEP11(2021)120}{{\em
  JHEP} {\bf 11} (2021)  120}, \href{http://arxiv.org/abs/2108.12998}{{\tt
  arXiv:2108.12998 [hep-ph]}}.

\bibitem{Jodlowski:2020vhr}
K.~Jod\l{}owski and S.~Trojanowski, ``{Neutrino beam-dump experiment with FASER
  at the LHC},'' \href{http://dx.doi.org/10.1007/JHEP05(2021)191}{{\em JHEP}
  {\bf 05} (2021)  191}, \href{http://arxiv.org/abs/2011.04751}{{\tt
  arXiv:2011.04751 [hep-ph]}}.

\bibitem{Ismail:2021dyp}
A.~Ismail, S.~Jana, and R.~M. Abraham, ``{Neutrino up-scattering via the dipole
  portal at forward LHC detectors},''
  \href{http://dx.doi.org/10.1103/PhysRevD.105.055008}{{\em Phys. Rev. D} {\bf
  105} (2022) no.~5, 055008}, \href{http://arxiv.org/abs/2109.05032}{{\tt
  arXiv:2109.05032 [hep-ph]}}.

\bibitem{Pospelov:2000bq}
M.~Pospelov and T.~ter Veldhuis, ``{Direct and indirect limits on the
  electromagnetic form-factors of WIMPs},''
  \href{http://dx.doi.org/10.1016/S0370-2693(00)00358-0}{{\em Phys. Lett. B}
  {\bf 480} (2000)  181--186}, \href{http://arxiv.org/abs/hep-ph/0003010}{{\tt
  arXiv:hep-ph/0003010}}.

\bibitem{Sigurdson:2004zp}
K.~Sigurdson, M.~Doran, A.~Kurylov, R.~R. Caldwell, and M.~Kamionkowski,
  ``{Dark-matter electric and magnetic dipole moments},''
  \href{http://dx.doi.org/10.1103/PhysRevD.70.083501}{{\em Phys. Rev. D} {\bf
  70} (2004)  083501}, \href{http://arxiv.org/abs/astro-ph/0406355}{{\tt
  arXiv:astro-ph/0406355}}. [Erratum: Phys.Rev.D 73, 089903 (2006)].

\bibitem{Schmidt:2012yg}
D.~Schmidt, T.~Schwetz, and T.~Toma, ``{Direct Detection of Leptophilic Dark
  Matter in a Model with Radiative Neutrino Masses},''
  \href{http://dx.doi.org/10.1103/PhysRevD.85.073009}{{\em Phys. Rev. D} {\bf
  85} (2012)  073009}, \href{http://arxiv.org/abs/1201.0906}{{\tt
  arXiv:1201.0906 [hep-ph]}}.

\bibitem{Ho:2012bg}
C.~M. Ho and R.~J. Scherrer, ``{Anapole Dark Matter},''
  \href{http://dx.doi.org/10.1016/j.physletb.2013.04.039}{{\em Phys. Lett. B}
  {\bf 722} (2013)  341--346}, \href{http://arxiv.org/abs/1211.0503}{{\tt
  arXiv:1211.0503 [hep-ph]}}.

\bibitem{Kopp:2014tsa}
J.~Kopp, L.~Michaels, and J.~Smirnov, ``{Loopy Constraints on Leptophilic Dark
  Matter and Internal Bremsstrahlung},''
  \href{http://dx.doi.org/10.1088/1475-7516/2014/04/022}{{\em JCAP} {\bf 04}
  (2014)  022}, \href{http://arxiv.org/abs/1401.6457}{{\tt arXiv:1401.6457
  [hep-ph]}}.

\bibitem{Ibarra:2015fqa}
A.~Ibarra and S.~Wild, ``{Dirac dark matter with a charged mediator: a
  comprehensive one-loop analysis of the direct detection phenomenology},''
  \href{http://dx.doi.org/10.1088/1475-7516/2015/05/047}{{\em JCAP} {\bf 05}
  (2015)  047}, \href{http://arxiv.org/abs/1503.03382}{{\tt arXiv:1503.03382
  [hep-ph]}}.

\bibitem{Sandick:2016zut}
P.~Sandick, K.~Sinha, and F.~Teng, ``{Simplified Dark Matter Models with
  Charged Mediators: Prospects for Direct Detection},''
  \href{http://dx.doi.org/10.1007/JHEP10(2016)018}{{\em JHEP} {\bf 10} (2016)
  018}, \href{http://arxiv.org/abs/1608.00642}{{\tt arXiv:1608.00642
  [hep-ph]}}.

\bibitem{Chu:2018qrm}
X.~Chu, J.~Pradler, and L.~Semmelrock, ``{Light dark states with
  electromagnetic form factors},''
  \href{http://dx.doi.org/10.1103/PhysRevD.99.015040}{{\em Phys. Rev. D} {\bf
  99} (2019) no.~1, 015040}, \href{http://arxiv.org/abs/1811.04095}{{\tt
  arXiv:1811.04095 [hep-ph]}}.

\bibitem{Trickle:2019ovy}
T.~Trickle, Z.~Zhang, and K.~M. Zurek, ``{Detecting Light Dark Matter with
  Magnons},'' \href{http://dx.doi.org/10.1103/PhysRevLett.124.201801}{{\em
  Phys. Rev. Lett.} {\bf 124} (2020) no.~20, 201801},
  \href{http://arxiv.org/abs/1905.13744}{{\tt arXiv:1905.13744 [hep-ph]}}.

\bibitem{Arina:2020mxo}
C.~Arina, A.~Cheek, K.~Mimasu, and L.~Pagani, ``{Light and Darkness:
  consistently coupling dark matter to photons via effective operators},''
  \href{http://dx.doi.org/10.1140/epjc/s10052-021-09010-1}{{\em Eur. Phys. J.
  C} {\bf 81} (2021) no.~3, 223}, \href{http://arxiv.org/abs/2005.12789}{{\tt
  arXiv:2005.12789 [hep-ph]}}.

\bibitem{Banerjee:2017hhz}
{\bf NA64} Collaboration, D.~Banerjee {\em et al.}, ``{Search for vector
  mediator of Dark Matter production in invisible decay mode},''
\href{http://arxiv.org/abs/1710.00971}{{\tt arXiv:1710.00971 [hep-ex]}}.

\bibitem{Akesson:2018vlm}
{\bf LDMX} Collaboration, T.~Åkesson {\em et al.}, ``{Light Dark Matter
  eXperiment (LDMX)},''
\href{http://arxiv.org/abs/1808.05219}{{\tt arXiv:1808.05219 [hep-ex]}}.

\bibitem{Battaglieri:2016ggd}
{\bf BDX} Collaboration, M.~Battaglieri {\em et al.}, ``{Dark Matter Search in
  a Beam-Dump eXperiment (BDX) at Jefferson Lab},''
\href{http://arxiv.org/abs/1607.01390}{{\tt arXiv:1607.01390 [hep-ex]}}.

\bibitem{Aubert:2001tu}
{\bf BaBar} Collaboration, B.~Aubert {\em et al.}, ``{The BaBar detector},''
  \href{http://dx.doi.org/10.1016/S0168-9002(01)02012-5}{{\em Nucl. Instrum.
  Meth.} {\bf A479} (2002)  1--116},
\href{http://arxiv.org/abs/hep-ex/0105044}{{\tt arXiv:hep-ex/0105044
  [hep-ex]}}.

\bibitem{Abe:2010gxa}
{\bf Belle-II} Collaboration, T.~Abe {\em et al.}, ``{Belle II Technical Design
  Report},''
\href{http://arxiv.org/abs/1011.0352}{{\tt arXiv:1011.0352 [physics.ins-det]}}.

\bibitem{Chu:2020ysb}
X.~Chu, J.-L. Kuo, and J.~Pradler, ``{Dark sector-photon interactions in
  proton-beam experiments},''
  \href{http://dx.doi.org/10.1103/PhysRevD.101.075035}{{\em Phys. Rev. D} {\bf
  101} (2020) no.~7, 075035}, \href{http://arxiv.org/abs/2001.06042}{{\tt
  arXiv:2001.06042 [hep-ph]}}.

\bibitem{Kuo:2021mtp}
J.-L. Kuo, M.~Pospelov, and J.~Pradler, ``{Terrestrial probes of
  electromagnetically interacting dark radiation},''
  \href{http://dx.doi.org/10.1103/PhysRevD.103.115030}{{\em Phys. Rev. D} {\bf
  103} (2021) no.~11, 115030}, \href{http://arxiv.org/abs/2102.08409}{{\tt
  arXiv:2102.08409 [hep-ph]}}.

\bibitem{Plestid:2020kdm}
R.~Plestid, V.~Takhistov, Y.-D. Tsai, T.~Bringmann, A.~Kusenko, and
  M.~Pospelov, ``{New Constraints on Millicharged Particles from Cosmic-ray
  Production},'' \href{http://dx.doi.org/10.1103/PhysRevD.102.115032}{{\em
  Phys. Rev. D} {\bf 102} (2020)  115032},
  \href{http://arxiv.org/abs/2002.11732}{{\tt arXiv:2002.11732 [hep-ph]}}.

\bibitem{ArguellesDelgado:2021lek}
C.~A. Arg\"uelles~Delgado, K.~J. Kelly, and V.~Mu\~noz Albornoz,
  ``{Millicharged particles from the heavens: single- and multiple-scattering
  signatures},'' \href{http://dx.doi.org/10.1007/JHEP11(2021)099}{{\em JHEP}
  {\bf 11} (2021)  099}, \href{http://arxiv.org/abs/2104.13924}{{\tt
  arXiv:2104.13924 [hep-ph]}}.

\bibitem{Chu:2019rok}
X.~Chu, J.-L. Kuo, J.~Pradler, and L.~Semmelrock, ``{Stellar probes of dark
  sector-photon interactions},''
  \href{http://dx.doi.org/10.1103/PhysRevD.100.083002}{{\em Phys. Rev. D} {\bf
  100} (2019) no.~8, 083002}, \href{http://arxiv.org/abs/1908.00553}{{\tt
  arXiv:1908.00553 [hep-ph]}}.

\bibitem{Chang:2019xva}
J.~H. Chang, R.~Essig, and A.~Reinert, ``{Light(ly)-coupled Dark Matter in the
  keV Range: Freeze-In and Constraints},''
  \href{http://dx.doi.org/10.1007/JHEP03(2021)141}{{\em JHEP} {\bf 03} (2021)
  141}, \href{http://arxiv.org/abs/1911.03389}{{\tt arXiv:1911.03389
  [hep-ph]}}.

\bibitem{Feng:2017uoz}
J.~L. Feng, I.~Galon, F.~Kling, and S.~Trojanowski, ``{ForwArd Search
  ExpeRiment at the LHC},''
  \href{http://dx.doi.org/10.1103/PhysRevD.97.035001}{{\em Phys. Rev. D} {\bf
  97} (2018) no.~3, 035001}, \href{http://arxiv.org/abs/1708.09389}{{\tt
  arXiv:1708.09389 [hep-ph]}}.

\bibitem{Feng:2017vli}
J.~L. Feng, I.~Galon, F.~Kling, and S.~Trojanowski, ``{Dark Higgs bosons at the
  ForwArd Search ExpeRiment},''
  \href{http://dx.doi.org/10.1103/PhysRevD.97.055034}{{\em Phys. Rev. D} {\bf
  97} (2018) no.~5, 055034}, \href{http://arxiv.org/abs/1710.09387}{{\tt
  arXiv:1710.09387 [hep-ph]}}.

\bibitem{Kling:2018wct}
F.~Kling and S.~Trojanowski, ``{Heavy Neutral Leptons at FASER},''
  \href{http://dx.doi.org/10.1103/PhysRevD.97.095016}{{\em Phys. Rev. D} {\bf
  97} (2018) no.~9, 095016}, \href{http://arxiv.org/abs/1801.08947}{{\tt
  arXiv:1801.08947 [hep-ph]}}.

\bibitem{Feng:2018noy}
J.~L. Feng, I.~Galon, F.~Kling, and S.~Trojanowski, ``{Axionlike particles at
  FASER: The LHC as a photon beam dump},''
  \href{http://dx.doi.org/10.1103/PhysRevD.98.055021}{{\em Phys. Rev.} {\bf
  D98} (2018) no.~5, 055021},
\href{http://arxiv.org/abs/1806.02348}{{\tt arXiv:1806.02348 [hep-ph]}}.

\bibitem{FASER:2021mtu}
{\bf FASER} Collaboration, H.~Abreu {\em et al.}, ``{First neutrino interaction
  candidates at the LHC},''
  \href{http://dx.doi.org/10.1103/PhysRevD.104.L091101}{{\em Phys. Rev. D} {\bf
  104} (2021) no.~9, L091101}, \href{http://arxiv.org/abs/2105.06197}{{\tt
  arXiv:2105.06197 [hep-ex]}}.

\bibitem{Ismail:2020yqc}
A.~Ismail, R.~Mammen~Abraham, and F.~Kling, ``{Neutral current neutrino
  interactions at FASER$\nu$},''
  \href{http://dx.doi.org/10.1103/PhysRevD.103.056014}{{\em Phys. Rev. D} {\bf
  103} (2021) no.~5, 056014}, \href{http://arxiv.org/abs/2012.10500}{{\tt
  arXiv:2012.10500 [hep-ph]}}.

\bibitem{Mosel:2022tqc}
U.~Mosel and K.~Gallmeister, ``{Neutrinos at FPF},''
  \href{http://arxiv.org/abs/2201.04008}{{\tt arXiv:2201.04008 [hep-ph]}}.

\bibitem{Kelly:2021mcd}
K.~J. Kelly, F.~Kling, D.~Tuckler, and Y.~Zhang, ``{Probing neutrino-portal
  dark matter at the Forward Physics Facility},''
  \href{http://dx.doi.org/10.1103/PhysRevD.105.075026}{{\em Phys. Rev. D} {\bf
  105} (2022) no.~7, 075026}, \href{http://arxiv.org/abs/2111.05868}{{\tt
  arXiv:2111.05868 [hep-ph]}}.

\bibitem{FASER:2018ceo}
{\bf FASER} Collaboration, A.~Ariga {\em et al.}, ``{Letter of Intent for
  FASER: ForwArd Search ExpeRiment at the LHC},''
  \href{http://arxiv.org/abs/1811.10243}{{\tt arXiv:1811.10243
  [physics.ins-det]}}.

\bibitem{FASER:2018eoc}
{\bf FASER} Collaboration, A.~Ariga {\em et al.}, ``{FASER\textquoteright{}s
  physics reach for long-lived particles},''
  \href{http://dx.doi.org/10.1103/PhysRevD.99.095011}{{\em Phys. Rev. D} {\bf
  99} (2019) no.~9, 095011}, \href{http://arxiv.org/abs/1811.12522}{{\tt
  arXiv:1811.12522 [hep-ph]}}.

\bibitem{FASER:2018bac}
{\bf FASER} Collaboration, A.~Ariga {\em et al.}, ``{Technical Proposal for
  FASER: ForwArd Search ExpeRiment at the LHC},''
  \href{http://arxiv.org/abs/1812.09139}{{\tt arXiv:1812.09139
  [physics.ins-det]}}.

\bibitem{FASER:2021cpr}
{\bf FASER} Collaboration, H.~Abreu {\em et al.}, ``{The trigger and data
  acquisition system of the FASER experiment},''
  \href{http://dx.doi.org/10.1088/1748-0221/16/12/P12028}{{\em JINST} {\bf 16}
  (2021) no.~12, P12028}, \href{http://arxiv.org/abs/2110.15186}{{\tt
  arXiv:2110.15186 [physics.ins-det]}}.

\bibitem{FASER:2021ljd}
{\bf FASER} Collaboration, H.~Abreu {\em et al.}, ``{The tracking detector of
  the FASER experiment},'' \href{http://arxiv.org/abs/2112.01116}{{\tt
  arXiv:2112.01116 [physics.ins-det]}}.

\bibitem{Boyd:2803084}
J.~Boyd, ``{The FASER W-Si High Precision Preshower Technical Proposal},''
  tech. rep., CERN, Geneva, Mar, 2022.
\newblock \url{https://cds.cern.ch/record/2803084}.

\bibitem{FASER:2019dxq}
{\bf FASER} Collaboration, H.~Abreu {\em et al.}, ``{Detecting and Studying
  High-Energy Collider Neutrinos with FASER at the LHC},''
  \href{http://dx.doi.org/10.1140/epjc/s10052-020-7631-5}{{\em Eur. Phys. J. C}
  {\bf 80} (2020) no.~1, 61}, \href{http://arxiv.org/abs/1908.02310}{{\tt
  arXiv:1908.02310 [hep-ex]}}.

\bibitem{FASER:2020gpr}
{\bf FASER} Collaboration, H.~Abreu {\em et al.}, ``{Technical Proposal:
  FASERnu},'' \href{http://arxiv.org/abs/2001.03073}{{\tt arXiv:2001.03073
  [physics.ins-det]}}.

\bibitem{SHiP:2020sos}
{\bf SHiP} Collaboration, C.~Ahdida {\em et al.}, ``{SND@LHC},''
  \href{http://arxiv.org/abs/2002.08722}{{\tt arXiv:2002.08722
  [physics.ins-det]}}.

\bibitem{Ahdida:2750060}
C.~Ahdida {\em et al.}, ``{SND@LHC - Scattering and Neutrino Detector at the
  LHC},'' tech. rep., CERN, Geneva, Jan, 2021.
\newblock \url{https://cds.cern.ch/record/2750060}.

\bibitem{MammenAbraham:2020hex}
R.~Mammen~Abraham {\em et al.}, ``{Forward Physics Facility - Snowmass 2021
  Letter of Interest},''.

\bibitem{Anchordoqui:2021ghd}
L.~A. Anchordoqui {\em et al.}, ``{The Forward Physics Facility: Sites,
  Experiments, and Physics Potential},''
  \href{http://arxiv.org/abs/2109.10905}{{\tt arXiv:2109.10905 [hep-ph]}}.

\bibitem{Feng:2022inv}
J.~L. Feng {\em et al.}, ``{The Forward Physics Facility at the High-Luminosity
  LHC},'' \href{http://arxiv.org/abs/2203.05090}{{\tt arXiv:2203.05090
  [hep-ex]}}.

\bibitem{Batell:2021blf}
B.~Batell, J.~L. Feng, and S.~Trojanowski, ``{Detecting Dark Matter with
  Far-Forward Emulsion and Liquid Argon Detectors at the LHC},''
  \href{http://dx.doi.org/10.1103/PhysRevD.103.075023}{{\em Phys. Rev. D} {\bf
  103} (2021) no.~7, 075023}, \href{http://arxiv.org/abs/2101.10338}{{\tt
  arXiv:2101.10338 [hep-ph]}}.

\bibitem{Foadi:2008qv}
R.~Foadi, M.~T. Frandsen, and F.~Sannino, ``{Technicolor Dark Matter},''
  \href{http://dx.doi.org/10.1103/PhysRevD.80.037702}{{\em Phys. Rev. D} {\bf
  80} (2009)  037702}, \href{http://arxiv.org/abs/0812.3406}{{\tt
  arXiv:0812.3406 [hep-ph]}}.

\bibitem{Bagnasco:1993st}
J.~Bagnasco, M.~Dine, and S.~D. Thomas, ``{Detecting technibaryon dark
  matter},'' \href{http://dx.doi.org/10.1016/0370-2693(94)90830-3}{{\em Phys.
  Lett. B} {\bf 320} (1994)  99--104},
  \href{http://arxiv.org/abs/hep-ph/9310290}{{\tt arXiv:hep-ph/9310290}}.

\bibitem{Antipin:2015xia}
O.~Antipin, M.~Redi, A.~Strumia, and E.~Vigiani, ``{Accidental Composite Dark
  Matter},'' \href{http://dx.doi.org/10.1007/JHEP07(2015)039}{{\em JHEP} {\bf
  07} (2015)  039}, \href{http://arxiv.org/abs/1503.08749}{{\tt
  arXiv:1503.08749 [hep-ph]}}.

\bibitem{Raby:1987ga}
S.~Raby and G.~West, ``{Experimental Consequences and Constraints for
  Magninos},'' \href{http://dx.doi.org/10.1016/0370-2693(87)90234-6}{{\em Phys.
  Lett. B} {\bf 194} (1987)  557--562}.

\bibitem{Pospelov:2008qx}
M.~Pospelov and A.~Ritz, ``{Resonant scattering and recombination of
  pseudo-degenerate WIMPs},''
  \href{http://dx.doi.org/10.1103/PhysRevD.78.055003}{{\em Phys. Rev. D} {\bf
  78} (2008)  055003}, \href{http://arxiv.org/abs/0803.2251}{{\tt
  arXiv:0803.2251 [hep-ph]}}.

\bibitem{Kling:2021fwx}
F.~Kling and S.~Trojanowski, ``{Forward experiment sensitivity estimator for
  the LHC and future hadron colliders},''
  \href{http://dx.doi.org/10.1103/PhysRevD.104.035012}{{\em Phys. Rev. D} {\bf
  104} (2021) no.~3, 035012}, \href{http://arxiv.org/abs/2105.07077}{{\tt
  arXiv:2105.07077 [hep-ph]}}.

\bibitem{Pierog:2013ria}
T.~Pierog, I.~Karpenko, J.~M. Katzy, E.~Yatsenko, and K.~Werner, ``{EPOS LHC:
  Test of collective hadronization with data measured at the CERN Large Hadron
  Collider},'' \href{http://dx.doi.org/10.1103/PhysRevC.92.034906}{{\em Phys.
  Rev.} {\bf C92} (2015) no.~3, 034906},
\href{http://arxiv.org/abs/1306.0121}{{\tt arXiv:1306.0121 [hep-ph]}}.

\bibitem{CRMC}
R.~Ulrich, T.~Pierog, and C.~Baus, ``Cosmic ray monte carlo package, crmc.''
  \url{https://doi.org/10.5281/zenodo.5270381}, Aug., 2021.

\bibitem{Sjostrand:2006za}
T.~Sjostrand, S.~Mrenna, and P.~Z. Skands, ``{PYTHIA 6.4 Physics and Manual},''
  \href{http://dx.doi.org/10.1088/1126-6708/2006/05/026}{{\em JHEP} {\bf 05}
  (2006)  026}, \href{http://arxiv.org/abs/hep-ph/0603175}{{\tt
  arXiv:hep-ph/0603175}}.

\bibitem{Sjostrand:2014zea}
T.~Sj\"ostrand, S.~Ask, J.~R. Christiansen, R.~Corke, N.~Desai, P.~Ilten,
  S.~Mrenna, S.~Prestel, C.~O. Rasmussen, and P.~Z. Skands, ``{An introduction
  to PYTHIA 8.2},'' \href{http://dx.doi.org/10.1016/j.cpc.2015.01.024}{{\em
  Comput. Phys. Commun.} {\bf 191} (2015)  159--177},
  \href{http://arxiv.org/abs/1410.3012}{{\tt arXiv:1410.3012 [hep-ph]}}.

\bibitem{Aaij:2015rla}
{\bf LHCb} Collaboration, R.~Aaij {\em et al.}, ``{Measurement of forward
  $J/\psi$ production cross-sections in $pp$ collisions at $\sqrt{s}=13$
  TeV},'' \href{http://dx.doi.org/10.1007/JHEP10(2015)172}{{\em JHEP} {\bf 10}
  (2015)  172}, \href{http://arxiv.org/abs/1509.00771}{{\tt arXiv:1509.00771
  [hep-ex]}}. [Erratum: JHEP 05, 063 (2017)].

\bibitem{Aaij:2018pfp}
{\bf LHCb} Collaboration, R.~Aaij {\em et al.}, ``{Measurement of $\Upsilon$
  production in $pp$ collisions at $\sqrt{s}$= 13 TeV},''
  \href{http://dx.doi.org/10.1007/JHEP07(2018)134}{{\em JHEP} {\bf 07} (2018)
  134}, \href{http://arxiv.org/abs/1804.09214}{{\tt arXiv:1804.09214
  [hep-ex]}}. [Erratum: JHEP 05, 076 (2019)].

\bibitem{Aaij:2019wfo}
{\bf LHCb} Collaboration, R.~Aaij {\em et al.}, ``{Measurement of $\psi$(2$S$)
  production cross-sections in proton-proton collisions at $\sqrt{s}$ = 7 and
  13 TeV},'' \href{http://dx.doi.org/10.1140/epjc/s10052-020-7638-y}{{\em Eur.
  Phys. J. C} {\bf 80} (2020) no.~3, 185},
  \href{http://arxiv.org/abs/1908.03099}{{\tt arXiv:1908.03099 [hep-ex]}}.

\bibitem{Alwall:2014hca}
J.~Alwall, R.~Frederix, S.~Frixione, V.~Hirschi, F.~Maltoni, O.~Mattelaer,
  H.~S. Shao, T.~Stelzer, P.~Torrielli, and M.~Zaro, ``{The automated
  computation of tree-level and next-to-leading order differential cross
  sections, and their matching to parton shower simulations},''
  \href{http://dx.doi.org/10.1007/JHEP07(2014)079}{{\em JHEP} {\bf 07} (2014)
  079}, \href{http://arxiv.org/abs/1405.0301}{{\tt arXiv:1405.0301 [hep-ph]}}.

\bibitem{Brem_precite}
S.~Foroughi-Abari and A.~Ritz. In preparation.

\bibitem{Batell:2021aja}
B.~Batell, J.~L. Feng, A.~Ismail, F.~Kling, R.~M. Abraham, and S.~Trojanowski,
  ``{Discovering dark matter at the LHC through its nuclear scattering in
  far-forward emulsion and liquid argon detectors},''
  \href{http://dx.doi.org/10.1103/PhysRevD.104.035036}{{\em Phys. Rev. D} {\bf
  104} (2021) no.~3, 035036}, \href{http://arxiv.org/abs/2107.00666}{{\tt
  arXiv:2107.00666 [hep-ph]}}.

\bibitem{Batell:2021snh}
B.~Batell, J.~L. Feng, M.~Fieg, A.~Ismail, F.~Kling, R.~M. Abraham, and
  S.~Trojanowski, ``{Hadrophilic dark sectors at the Forward Physics
  Facility},'' \href{http://dx.doi.org/10.1103/PhysRevD.105.075001}{{\em Phys.
  Rev. D} {\bf 105} (2022) no.~7, 075001},
  \href{http://arxiv.org/abs/2111.10343}{{\tt arXiv:2111.10343 [hep-ph]}}.

\bibitem{DUNE:2015lol}
{\bf DUNE} Collaboration, R.~Acciarri {\em et al.}, ``{Long-Baseline Neutrino
  Facility (LBNF) and Deep Underground Neutrino Experiment (DUNE)}: {Conceptual
  Design Report, Volume 2: The Physics Program for DUNE at LBNF},''
  \href{http://arxiv.org/abs/1512.06148}{{\tt arXiv:1512.06148
  [physics.ins-det]}}.

\bibitem{Bai:2020ukz}
W.~Bai, M.~Diwan, M.~V. Garzelli, Y.~S. Jeong, and M.~H. Reno, ``{Far-forward
  neutrinos at the Large Hadron Collider},''
  \href{http://dx.doi.org/10.1007/JHEP06(2020)032}{{\em JHEP} {\bf 06} (2020)
  032}, \href{http://arxiv.org/abs/2002.03012}{{\tt arXiv:2002.03012
  [hep-ph]}}.

\bibitem{Kling:2021gos}
F.~Kling and L.~J. Nevay, ``{Forward neutrino fluxes at the LHC},''
  \href{http://dx.doi.org/10.1103/PhysRevD.104.113008}{{\em Phys. Rev. D} {\bf
  104} (2021) no.~11, 113008}, \href{http://arxiv.org/abs/2105.08270}{{\tt
  arXiv:2105.08270 [hep-ph]}}.

\bibitem{Bai:2021ira}
W.~Bai, M.~Diwan, M.~V. Garzelli, Y.~S. Jeong, F.~K. Kumar, and M.~H. Reno,
  ``{Parton distribution function uncertainties in theoretical predictions for
  far-forward tau neutrinos at the Large Hadron Collider},''
  \href{http://arxiv.org/abs/2112.11605}{{\tt arXiv:2112.11605 [hep-ph]}}.

\bibitem{Bai:2022jcs}
W.~Bai, M.~V. Diwan, M.~V. Garzelli, Y.~S. Jeong, K.~Kumar, and M.~H. Reno,
  ``{Prompt electron and tau neutrinos and antineutrinos in the forward region
  at the LHC},'' in {\em {2022 Snowmass Summer Study}}.
\newblock 3, 2022.
\newblock \href{http://arxiv.org/abs/2203.07212}{{\tt arXiv:2203.07212
  [hep-ph]}}.

\bibitem{Andreopoulos:2009rq}
C.~Andreopoulos {\em et al.}, ``{The GENIE Neutrino Monte Carlo Generator},''
  \href{http://dx.doi.org/10.1016/j.nima.2009.12.009}{{\em Nucl. Instrum. Meth.
  A} {\bf 614} (2010)  87--104}, \href{http://arxiv.org/abs/0905.2517}{{\tt
  arXiv:0905.2517 [hep-ph]}}.

\bibitem{Andreopoulos:2015wxa}
C.~Andreopoulos, C.~Barry, S.~Dytman, H.~Gallagher, T.~Golan, R.~Hatcher,
  G.~Perdue, and J.~Yarba, ``{The GENIE Neutrino Monte Carlo Generator: Physics
  and User Manual},'' \href{http://arxiv.org/abs/1510.05494}{{\tt
  arXiv:1510.05494 [hep-ph]}}.

\bibitem{ArgoNeuT:2018tvi}
{\bf ArgoNeuT} Collaboration, R.~Acciarri {\em et al.}, ``{Demonstration of
  MeV-Scale Physics in Liquid Argon Time Projection Chambers Using ArgoNeuT},''
  \href{http://dx.doi.org/10.1103/PhysRevD.99.012002}{{\em Phys. Rev. D} {\bf
  99} (2019) no.~1, 012002}, \href{http://arxiv.org/abs/1810.06502}{{\tt
  arXiv:1810.06502 [hep-ex]}}.

\bibitem{Choi:2020mbk}
S.~Choi {\em et al.}, ``{Letter of Intent: Search for sub-millicharged
  particles at J-PARC},'' \href{http://arxiv.org/abs/2007.06329}{{\tt
  arXiv:2007.06329 [physics.ins-det]}}.

\bibitem{Kim:2021eix}
J.~H. Kim, I.~S. Hwang, and J.~H. Yoo, ``{Search for sub-millicharged particles
  at J-PARC},'' \href{http://dx.doi.org/10.1007/JHEP05(2021)031}{{\em JHEP}
  {\bf 05} (2021)  031}, \href{http://arxiv.org/abs/2102.11493}{{\tt
  arXiv:2102.11493 [hep-ex]}}.

\bibitem{Erickcek:2007jv}
A.~L. Erickcek, P.~J. Steinhardt, D.~McCammon, and P.~C. McGuire,
  ``{Constraints on the Interactions between Dark Matter and Baryons from the
  X-ray Quantum Calorimetry Experiment},''
  \href{http://dx.doi.org/10.1103/PhysRevD.76.042007}{{\em Phys. Rev.} {\bf
  D76} (2007)  042007},
\href{http://arxiv.org/abs/0704.0794}{{\tt arXiv:0704.0794 [astro-ph]}}.

\bibitem{Dubovsky:2003yn}
S.~L. Dubovsky, D.~S. Gorbunov, and G.~I. Rubtsov, ``{Narrowing the window for
  millicharged particles by CMB anisotropy},''
  \href{http://dx.doi.org/10.1134/1.1675909}{{\em JETP Lett.} {\bf 79} (2004)
  1--5}, \href{http://arxiv.org/abs/hep-ph/0311189}{{\tt arXiv:hep-ph/0311189
  [hep-ph]}}.
[Pisma Zh. Eksp. Teor. Fiz.79,3(2004)].

\bibitem{Dolgov:2013una}
A.~D. Dolgov, S.~L. Dubovsky, G.~I. Rubtsov, and I.~I. Tkachev, ``{Constraints
  on millicharged particles from Planck data},''
  \href{http://dx.doi.org/10.1103/PhysRevD.88.117701}{{\em Phys. Rev.} {\bf
  D88} (2013) no.~11, 117701},
\href{http://arxiv.org/abs/1310.2376}{{\tt arXiv:1310.2376 [hep-ph]}}.

\bibitem{McCammon:2002gb}
D.~McCammon {\em et al.}, ``{A High spectral resolution observation of the soft
  x-ray diffuse background with thermal detectors},''
  \href{http://dx.doi.org/10.1086/341727}{{\em Astrophys. J.} {\bf 576} (2002)
  188--203}, \href{http://arxiv.org/abs/astro-ph/0205012}{{\tt
  arXiv:astro-ph/0205012}}.

\bibitem{CHARM-II:1989nic}
{\bf CHARM-II} Collaboration, K.~De~Winter {\em et al.}, ``{A Detector for the
  Study of Neutrino - Electron Scattering},''
  \href{http://dx.doi.org/10.1016/0168-9002(89)91190-X}{{\em Nucl. Instrum.
  Meth. A} {\bf 278} (1989)  670}.

\bibitem{CHARM-II:1994dzw}
{\bf CHARM-II} Collaboration, P.~Vilain {\em et al.}, ``{Precision measurement
  of electroweak parameters from the scattering of muon-neutrinos on
  electrons},'' \href{http://dx.doi.org/10.1016/0370-2693(94)91421-4}{{\em
  Phys. Lett. B} {\bf 335} (1994)  246--252}.

\bibitem{CMS:2017zts}
{\bf CMS} Collaboration, A.~M. Sirunyan {\em et al.}, ``{Search for new physics
  in final states with an energetic jet or a hadronically decaying $W$ or $Z$
  boson and transverse momentum imbalance at $\sqrt{s}=13\text{ }\text{
  }\mathrm{TeV}$},'' \href{http://dx.doi.org/10.1103/PhysRevD.97.092005}{{\em
  Phys. Rev. D} {\bf 97} (2018) no.~9, 092005},
  \href{http://arxiv.org/abs/1712.02345}{{\tt arXiv:1712.02345 [hep-ex]}}.

\bibitem{BaBar:2001yhh}
{\bf BaBar} Collaboration, B.~Aubert {\em et al.}, ``{The BaBar detector},''
  \href{http://dx.doi.org/10.1016/S0168-9002(01)02012-5}{{\em Nucl. Instrum.
  Meth. A} {\bf 479} (2002)  1--116},
  \href{http://arxiv.org/abs/hep-ex/0105044}{{\tt arXiv:hep-ex/0105044}}.

\bibitem{Ball:1980ojt}
R.~Ball {\em et al.}, ``{The neutrino beam dump experiment at Fermilab
  (E613)},'' {\em eConf} {\bf C801002} (1980)  172--174.

\bibitem{LSND:1996jxj}
{\bf LSND} Collaboration, C.~Athanassopoulos {\em et al.}, ``{The Liquid
  scintillator neutrino detector and LAMPF neutrino source},''
  \href{http://dx.doi.org/10.1016/S0168-9002(96)01155-2}{{\em Nucl. Instrum.
  Meth. A} {\bf 388} (1997)  149--172},
  \href{http://arxiv.org/abs/nucl-ex/9605002}{{\tt arXiv:nucl-ex/9605002}}.

\bibitem{MiniBooNEDM:2018cxm}
{\bf MiniBooNE DM} Collaboration, A.~A. Aguilar-Arevalo {\em et al.}, ``{Dark
  Matter Search in Nucleon, Pion, and Electron Channels from a Proton Beam Dump
  with MiniBooNE},'' \href{http://dx.doi.org/10.1103/PhysRevD.98.112004}{{\em
  Phys. Rev. D} {\bf 98} (2018) no.~11, 112004},
  \href{http://arxiv.org/abs/1807.06137}{{\tt arXiv:1807.06137 [hep-ex]}}.

\bibitem{NA64:2017vtt}
{\bf NA64} Collaboration, D.~Banerjee {\em et al.}, ``{Search for vector
  mediator of Dark Matter production in invisible decay mode},''
  \href{http://dx.doi.org/10.1103/PhysRevD.97.072002}{{\em Phys. Rev. D} {\bf
  97} (2018) no.~7, 072002}, \href{http://arxiv.org/abs/1710.00971}{{\tt
  arXiv:1710.00971 [hep-ex]}}.

\bibitem{L3:2003yon}
{\bf L3} Collaboration, P.~Achard {\em et al.}, ``{Single photon and
  multiphoton events with missing energy in $e^{+} e^{-}$ collisions at LEP},''
  \href{http://dx.doi.org/10.1016/j.physletb.2004.01.010}{{\em Phys. Lett. B}
  {\bf 587} (2004)  16--32}, \href{http://arxiv.org/abs/hep-ex/0402002}{{\tt
  arXiv:hep-ex/0402002}}.

\bibitem{Fortin:2011hv}
J.-F. Fortin and T.~M.~P. Tait, ``{Collider Constraints on Dipole-Interacting
  Dark Matter},'' \href{http://dx.doi.org/10.1103/PhysRevD.85.063506}{{\em
  Phys. Rev. D} {\bf 85} (2012)  063506},
  \href{http://arxiv.org/abs/1103.3289}{{\tt arXiv:1103.3289 [hep-ph]}}.

\bibitem{Jodlowski:2019ycu}
K.~Jod\l{}owski, F.~Kling, L.~Roszkowski, and S.~Trojanowski, ``{Extending the
  reach of FASER, MATHUSLA, and SHiP towards smaller lifetimes using secondary
  particle production},''
  \href{http://dx.doi.org/10.1103/PhysRevD.101.095020}{{\em Phys. Rev. D} {\bf
  101} (2020) no.~9, 095020}, \href{http://arxiv.org/abs/1911.11346}{{\tt
  arXiv:1911.11346 [hep-ph]}}.

\bibitem{Rodrigues:2019nct}
E.~Rodrigues, ``{The Scikit-HEP Project},''
  \href{http://dx.doi.org/10.1051/epjconf/201921406005}{{\em EPJ Web Conf.}
  {\bf 214} (2019)  06005}, \href{http://arxiv.org/abs/1905.00002}{{\tt
  arXiv:1905.00002 [physics.comp-ph]}}.

\bibitem{Rodrigues:2020syo}
E.~Rodrigues {\em et al.}, ``{The Scikit HEP Project -- overview and
  prospects},'' \href{http://dx.doi.org/10.1051/epjconf/202024506028}{{\em EPJ
  Web Conf.} {\bf 245} (2020)  06028},
  \href{http://arxiv.org/abs/2007.03577}{{\tt arXiv:2007.03577
  [physics.comp-ph]}}.

\bibitem{pylhe}
L.~Heinrich and M.~Feickert, ``{pylhe: v0.2.1}.''
  \url{https://github.com/scikit-hep/pylhe}.

\bibitem{henry_schreiner_2022_5942083}
H.~Schreiner, J.~Pivarski, E.~Rodrigues, F.~Bruggisser, L.~Kreczko, N!no,
  P.~Fackeldey, S.~Pérez, and T.~G. Badger, ``scikit-hep/vector: Version
  0.8.5.'' \url{https://doi.org/10.5281/zenodo.5942083}, Feb., 2022.

\bibitem{ParticleDataGroup:2020ssz}
{\bf Particle Data Group} Collaboration, P.~A. Zyla {\em et al.}, ``{Review of
  Particle Physics},'' \href{http://dx.doi.org/10.1093/ptep/ptaa104}{{\em PTEP}
  {\bf 2020} (2020) no.~8, 083C01}.

\end{thebibliography}\endgroup

\end{document}